\documentclass[aps,prb,superscriptaddress,reprint]{revtex4-1}
\usepackage[english]{babel}
\usepackage{amsmath,amsthm,amssymb}
\usepackage{amsfonts}
\usepackage{xspace}
\usepackage{bm}
\usepackage{graphicx}

\usepackage[separate-uncertainty = true, exponent-product = \times]{siunitx}
\usepackage{hyperref}
\usepackage{dcolumn}

\begin{document}
\newcommand{\la}{\langle}
\newcommand{\ra}{\rangle}
\newcommand{\eq}[1]{\begin{equation}#1\end{equation}}
\newcommand{\multeq}[1]{\begin{equation}\begin{split}#1\end{split}\end{equation}}
\newcommand{\p}{\partial}
\newcommand{\pfrac}[2]{\frac{\p#1}{\p#2}}
\renewcommand{\dfrac}[2]{\frac{d#1}{d#2}}
\newcommand{\mat}[1]{\left(\begin{smallmatrix}#1\end{smallmatrix}\right)}

\newcommand{\dT}{\Delta T_\mathrm{me}}
\newcommand{\gsm}{g_\mathrm{r}}

\renewcommand{\figurename}{FIG.}
\renewcommand{\tablename}{TABLE}

\title{Magnon, phonon and electron temperature profiles and the spin Seebeck effect in magnetic insulator/normal metal hybrid structures}%

\author{Michael Schreier}
\email{michael.schreier@wmi.badw.de}
\affiliation{Walther-Mei\ss ner-Institut, Bayerische Akademie der Wissenschaften, Garching, Germany}

\author{Akashdeep Kamra}
\affiliation{Kavli Institute of Nanoscience, Delft University of Technology, Delft, The Netherlands}

\author{Mathias Weiler}
\affiliation{Walther-Mei\ss ner-Institut, Bayerische Akademie der Wissenschaften, Garching, Germany}
\affiliation{Present address: National Institute of Standards and Technology, Boulder, CO, USA}

\author{Jiang Xiao}
\affiliation{Department of Physics and State Key Laboratory of Surface Physics, Fudan University, Shanghai, China}

\author{Gerrit E.~W. Bauer}
\affiliation{Kavli Institute of Nanoscience, Delft University of Technology, Delft, The Netherlands}
\affiliation{Institute for Materials Research and WPI-AIMR, Tohoku University, Sendai, Japan}

\author{Rudolf Gross}
\affiliation{Walther-Mei\ss ner-Institut, Bayerische Akademie der Wissenschaften, Garching, Germany}
\affiliation{Physik Department, Technische Universit\"at M\"unchen, Garching, Germany}

\author{Sebastian T.~B. Goennenwein}
\affiliation{Walther-Mei\ss ner-Institut, Bayerische Akademie der Wissenschaften, Garching, Germany}
\date{\today}%
\begin{abstract}
We calculate the phonon, electron and magnon temperature profiles in yttrium iron garnet/platinum bilayers by diffusive theory with appropriate boundary conditions, in particular taking into account interfacial thermal resistances. Our calculations show that in thin film hybrids, the interface magnetic heat conductance qualitatively affects the magnon temperature. Based on published material parameters we assess the degree of non-equilibrium at the yttrium iron garnet/platinum interface. The magnitude of the spin Seebeck effect derived from this approach compares well with experimental results for the longitudinal spin Seebeck effect. Additionally we address the temperature profiles in the transverse spin Seebeck effect.
\end{abstract}
\maketitle

\section{Introduction}\label{sec:intro}
\begin{figure}%
\includegraphics[width=\columnwidth]{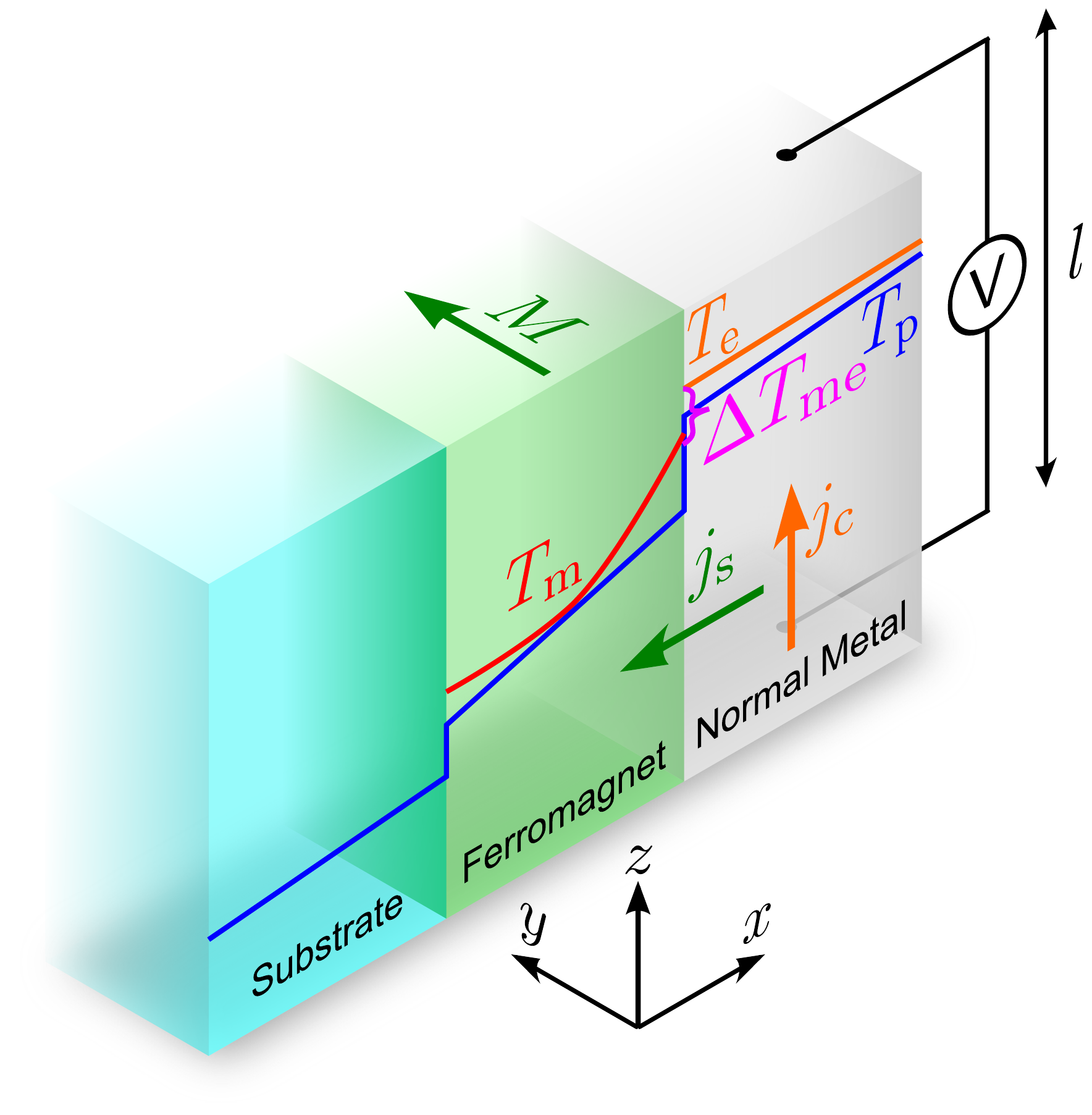}%
\caption{(Color online) In the (longitudinal) spin Seebeck effect a difference $\dT$ between the temperature $T_\mathrm{m}$ of the magnons in a ferromagnet (FM) and the temperature $T_\mathrm{e}$ of the electrons in a normal metal (NM) causes a spin current $j_\mathrm{s}$ between the ferromagnet and the normal metal that is converted into a charge current $\pmb j_\mathrm{c}\propto j_\mathrm{s}\hat{\pmb{x}}\times\hat{\pmb{s}}$ by the inverse spin Hall effect, where $\hat{\pmb{s}}=-\pmb M/|\pmb M|$ is the polarization vector of the spin current and $\pmb M$ is the magnetization vector.}%
\label{fig:thSSE}%
\end{figure}

The spin Seebeck effect~\cite{Uchida2008,Uchida2010} (SSE), a recent addition to the field of spin caloritronics,~\cite{Bauer2012} allows us to thermally generate pure spin currents. While the spin Seebeck effect itself has been experimentally established,~\cite{Uchida2008,Uchida2010,Uchida2010a,uchida2010b,Jaworski2010,Weiler2012,Meier2013,Qu2013,Kikkawa2013} an agreement between experiments and theory~\cite{Xiao2010,*Xiao2010a,Adachi2011} has proven elusive. In experimental publications the average \emph{temperature gradient} across the entire sample is usually quoted but the thermodynamic state at the interface at which the spin current is generated, could not be measured yet. However, for comparison with theory, the knowledge of the actual \emph{temperature difference} $\dT$ between the magnon and the electron systems at the  ferromagnet/normal metal interface is crucial, since it drives the spin Seebeck effect.~\cite{Xiao2010} The temperature difference $\dT$ arises due to different thermal properties and boundary conditions for the magnons, phonons and electrons in the ferromagnet/normal metal hybrids used in experiments. The phonon ($T_\mathrm{p}$), electron ($T_\mathrm{e}$) and magnon ($T_\mathrm{m}$) temperature profiles in a substrate/ferromagnet/normal metal multilayer structure are schematically depicted in Fig.~\ref{fig:thSSE}. As detailed in this paper, the temperature profiles can show discontinuities at the material interfaces due to interface properties such as the Kapitza resistance.~\cite{Kapitza1941} Temperature profiles are not easily measurable for a non-equilibrium situation in which magnon, phonon, and electron temperatures differ. An in depth analysis and interpretation of experimental spin Seebeck effect data is to date possible only by modeling the magnon, phonon, and electron temperature profiles based on the relevant material parameters. Especially for magnetic insulators the determination of the phonon temperature $T_\mathrm{p}$ profile is of central importance in this approach.~\cite{Kaganov1957,Sanders1977}\\
In this paper, we model and calculate the phonon, electron and magnon temperature profiles in ferromagnet/normal metal hybrid structures, by solving the heat transport equations with appropriate boundary conditions. In particular, we explicitly take into account the heat current carried by the pumped spin current through the interface. For the sake of simplicity, we limit the discussion to hybrids based on the ferrimagnetic insulator yttrium iron garnet (Y$_3$Fe$_5$O$_{12}$, YIG). However, our approach can easily be extended to other material systems. We analytically solve the heat transport equations when possible, and use numerical simulations via three-dimensional finite element (3D FEM) solvers (COMSOL~\footnote{COMSOL Multiphysics\textsuperscript{\textregistered} 4.3a}) otherwise. The magnon temperature calculations are based on the work by Sanders and Walton~\cite{Sanders1977} and Xiao \textit{et al.},~\cite{Xiao2010} that we extensively use with a focus on ultrathin films in which interfacial effects become important for the thermal transport. The $T_\mathrm{p}$, $T_\mathrm{e}$, and $T_\mathrm{m}$ profiles thus obtained yield the temperature difference $\dT$ at the magnetic insulator/normal metal interfaces, from which the spin Seebeck voltage can be inferred.~\cite{Xiao2010} The voltages thus calculated agree well with our spatially resolved longitudinal spin Seebeck effect measurements.~\cite{Weiler2012,Schreier2012} We also apply our calculations to the transverse spin Seebeck configuration and find phonon and magnon temperatures profiles that are in good agreement with the experimental results by Agrawal \textit{et al.}.~\cite{Agrawal2012} We finally illustrate the role of out of plane thermal gradients for the transverse spin Seebeck effect.\\
The paper is organized as follows: We first start with a brief overview over the experimental technique we used for the study of YIG/Pt thin film bilayers (Sec.~\ref{sec:experiment}), followed by an introduction to the theory of the spin Seebeck effect as proposed by Xiao \textit{et al.}~\cite{Xiao2010} (Sec.~\ref{sec:TheoSSE}). We then develop the model for the coupled phonon, electron, and magnon heat transport including interfacial thermal resistances (Secs.~\ref{sec:phonheat} and~\ref{sec:analytic}). We study the analytical solution of the coupled system in multilayers with lateral translational symmetry (Sec.~\ref{sec:numeric}) and extend our findings to full 3D numerical simulations of the thermal profiles found in our experiments~\cite{Weiler2012,Schreier2012} (Sec.~\ref{sec:thermlandsc}). Finally we apply our model to the transverse spin Seebeck effect and compare it to recent experiments~\cite{Agrawal2012,Uchida2010} (Sec.~\ref{sec:transSSE}).

\section{Experiment}\label{sec:experiment}
The samples in our experiments~\cite{Weiler2012,Schreier2012} consist of a series of thin YIG films grown on $\SI{500}{\micro m}$ gadolinium gallium garnet (Gd$_3$Ga$_5$O$_{12}$, GGG) and yttrium aluminium garnet (Y$_3$Al$_5$O$_{12}$, YAG) substrates via pulsed laser deposition with thicknesses in the range of $40-\SI{70}{\nano m}$. On top of the YIG layer, thin ($1.5-\SI{20}{\nano m}$) platinum or gold films were deposited \textit{in situ}, without breaking the vacuum, using electron beam evaporation. The fabrication of the samples is described in more detail in Refs.~\onlinecite{Geprags2012,Althammer2013}. 
After the growth process, the samples were patterned into Hall bar mesa structures (width $w=\SI{80}{\micro m}$, length $l=\SI{950}{\micro m}$) using optical lithography and Argon ion beam milling, mounted in a chip carrier system, and inserted into a home-built 2D vector magnet. We then use a scanning focused, laser beam to locally heat the samples.~\cite{Weiler2012} The heating by the laser ($\lambda_\mathrm{Laser}=\SI{660}{\nano m}$) generates a thermal gradient normal to the sample plane and hence a spin current via the spin Seebeck effect. The spin current is converted into a charge current via the inverse spin Hall effect in the platinum, and can thus be detected as a voltage using conventional electronics. For laser powers of $\SI{1}{\milli W}\leq P\leq\SI{60}{\milli W}$ we detect voltages in the range of $\SI{10}{\nano V}\lesssim V_\mathrm{SSE}\lesssim\SI{10}{\micro V}$.  All spin Seebeck effect experiments were performed at room temperature. The measured voltages are entirely attributed to the spin Seebeck effect since our platinum layers do not show any significant static proximity polarization~\cite{Geprags2012} that could create contributions from the anomalous Nernst effect. This conclusion is supported by recent studies~\cite{Kikkawa2013} that report spin Seebeck effect in YIG/Pt heterostructures far in excess of any possible contributions from the anomalous Nernst effect.\\

\section{Theory of the spin Seebeck effect}\label{sec:TheoSSE}
According to Xiao \textit{et al.}, the spin Seebeck voltage is given by the following equation~\cite{Xiao2010}
\begin{equation}
	V_\mathrm{SSE}=\frac{\gsm \gamma \hbar k_\mathrm{B}}{2\pi M_\mathrm{s} V_\mathrm{a}}\dT\cdot\frac{2e}{\hbar}\theta_\mathrm{H}\rho l\cdot \eta\cdot\frac{\lambda}{t}\tanh{\left(\frac{t}{2\lambda}\right)},\label{eq:SSE}
\end{equation}
where we assume~\footnote{An expression for the spin backflow $\eta$ in the spin Seebeck effect has not been established yet. We use the expression for spin pumping here due to the closely related physics governing the two phenomena.} a backflow correction factor~\cite{Jiao2013,Chen2013a} from spin diffusion theory in the normal metal of
\begin{equation}
	 \eta=\left[1+2\gsm\rho\lambda\frac{e^2}{h}\coth{\left(\frac{t}{\lambda}\right)}\right]^{-1}.\label{eq:backflow}
\end{equation}
Here $\theta_\mathrm{H}$ is the spin Hall angle,~\cite{Takahashi2008} $\rho$ is the electrical resistivity of the sample, $l$ is the length of the sample (the distance between the voltage contacts determining the voltage $V_\mathrm{SSE}=E_\mathrm{SSE}\cdot l$ transverse to the magnetization orientation of the ferromagnet), $\gsm=\mathrm{Re}\left(g^{\uparrow\!\downarrow}\right)$ is the real part of the spin mixing interface conductance per unit area,~\cite{Brataas2002} $\gamma=g \frac{e}{2m}$ is the gyromagnetic ratio with $g$ as the Land\'{e} $g$-factor and the electron mass $m$, $e=\left|e\right|$ is the elementary charge, $k_\mathrm{B}$ is the Boltzmann constant, $h$ is the Planck constant, $M_\mathrm{s}$ is the saturation magnetization of the ferromagnet, $\dT=T_\mathrm{m}-T_\mathrm{e}$ the temperature difference between the magnons in the ferromagnet and the electrons in the normal metal at the ferromagnet/normal metal interface, $\lambda$ is the spin diffusion length in the normal metal, and $t$ is the thickness of the normal metal film. $V_\mathrm{a}$ is the magnetic coherence volume given by~\cite{Xiao2010a}
\begin{equation}
	V_\mathrm{a}=\frac{2}{3\zeta(5/2)}\left(\frac{4\pi D}{k_\mathrm{B}T}\right)^{3/2},
\label{eq:Va}
\end{equation}
where $\zeta$ is the Riemann Zeta function and $D$ is the spin wave stiffness.

As evident from Eq.~\eqref{eq:SSE}, the spin Seebeck voltage hinges on $\dT$. In the following, we therefore discuss the evaluation of $T_\mathrm{p}$, $T_\mathrm{e}$ and $T_\mathrm{m}$ in thin film and bulk-like heterostructures. From these temperature profiles one can then quantitatively infer $\dT$ and thus calculate the spin Seebeck voltage.

\section{Phonon heat transport}~\label{sec:phonheat}
Heat transport in a homogeneous material with a single heat carrier (e.g.,~phonons) is described by the heat diffusion equation~\cite{Fourier1822}
\begin{equation}
	\nabla^2T-\frac{1}{k}\pfrac{T}{t}=-\frac{Q}{\kappa},
\label{eq:simpleheat}
\end{equation}
where $Q$ is the heating power density, $\kappa$ is the thermal conductivity, and $k = \kappa/\varrho C$ is the thermal diffusivity, with $\varrho$ as the mass density and $C$ as the heat capacity of the material. For simplicity, we assume $\kappa$, $\varrho$, and $C$ to be spatially homogeneous and temperature independent. The latter assumption is valid as long as the considered temperature changes are small. In a heterostructure consisting of several layers stacked on top of one another, Eq.~\eqref{eq:simpleheat} has to be solved piecewise for each layer~\cite{Reichling1994}:
\begin{equation}
	\nabla^2T_i-\frac{1}{k_i}\pfrac{T_i}{t}=-\frac{Q_i}{\kappa_i},\label{eq:coupledheat}
\end{equation}
with boundary conditions for the temperatures $T_i$ and $T_{i+1}$ on both sides of an interface
\begin{equation}
	\begin{split}
		-\kappa_i\left.\pfrac{T_i}{x}\right|_\mathrm{interface} &= \frac{1}{R_{\mathrm{th},i}}\left.[T_i-T_{i+1}]\right|_\mathrm{interface},\\
		-\kappa_{i+1}\left.\pfrac{T_{i+1}}{x}\right|_\mathrm{interface} &= \frac{1}{R_{\mathrm{th},i}}\left.[T_i-T_{i+1}]\right|_\mathrm{interface},\label{eq:thermbound}
		\end{split}
\end{equation}
where $i$ is the index for the individual materials (or layers in our case, i.e. the normal metal, the ferromagnet or the substrate), and $R_{\mathrm{th},i}$ is the interfacial thermal resistance between layer $i$ and $i+1$. In the steady state Eq.~\eqref{eq:coupledheat} reduces to
\begin{equation}
	\nabla^2T_i=-\frac{Q_i}{\kappa_i}.
\label{eq:heatSS}
\end{equation}
Solving Eq.~\eqref{eq:heatSS} together with the appropriate boundary conditions [Eq.~\eqref{eq:thermbound}] leads to the (phonon) temperature distribution.\\
In the samples in question, however, the heat is not carried exclusively by phonons, but by electrons and magnons as well. To draw a complete picture of the arising temperature profiles one therefore has to take the coupling between the individual systems into account. While both thermal magnons and electrons have relatively short interaction times with phonons,~\cite{Kumar1967,Caffrey2005,Lin2008} in our few nanometer thick films, equilibration between the individual systems might be incomplete. We therefore simulate our experiments by explicitly including phonons, electrons and magnons separately as outlined in Sec.~\ref{sec:analytic}.

\section{Coupled magnonic and phononic heat transport}\label{sec:analytic}
\begin{figure}%
\includegraphics[width=\columnwidth]{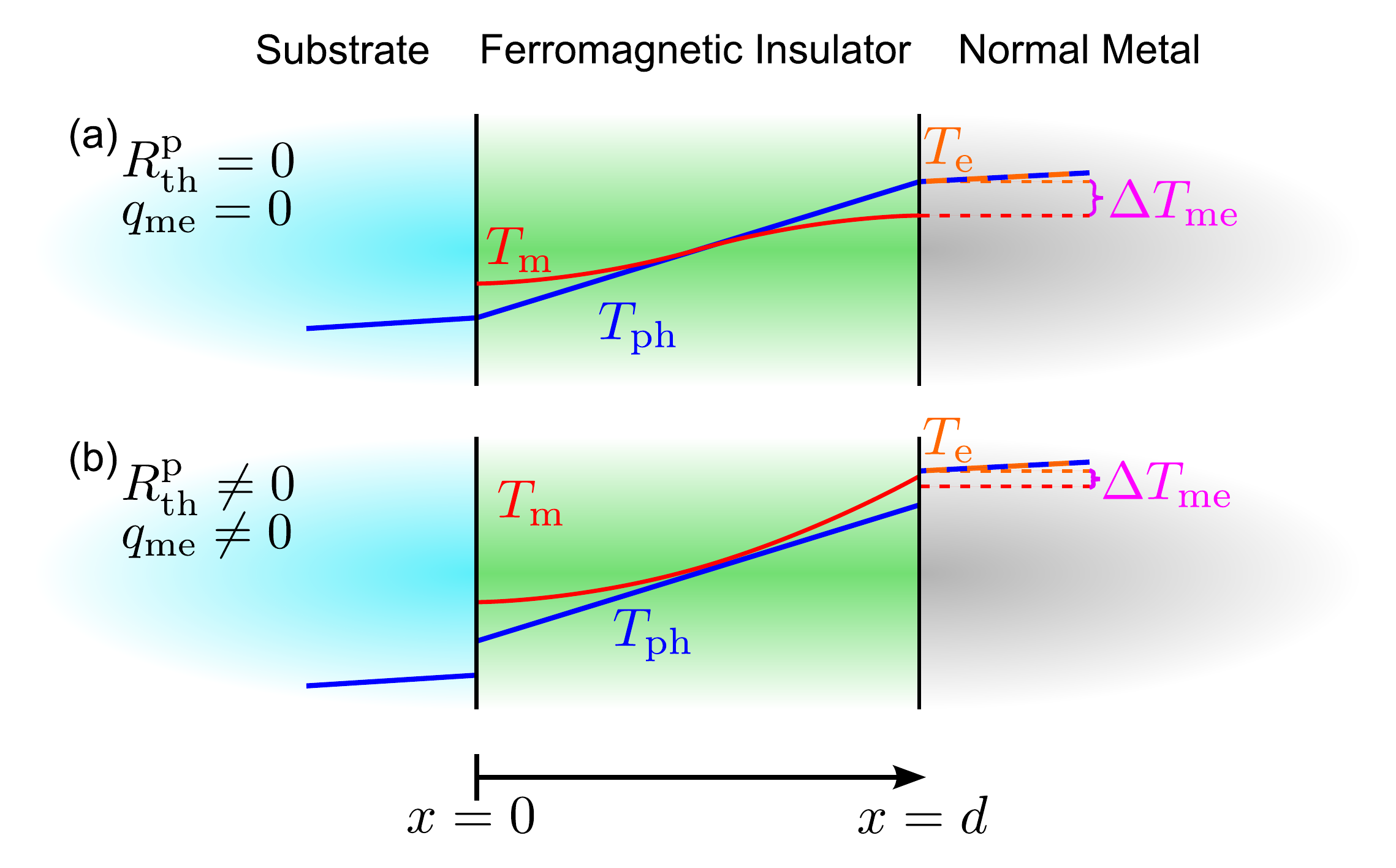}%
\caption{(Color online) Schematic phonon and magnon temperature profiles in a layered structure. We here assume identical phonon and electron temperatures in the normal metal. 
(a) For vanishing interfacial thermal resistance $R_\mathrm{th}=0$ and no
spin mediated heat current across the ferromagnet/normal metal interface,~\cite{Xiao2010} the temperature difference $\dT=T_\mathrm{m}-T_\mathrm{e}=T_\mathrm{m}-T_\mathrm{p}$ depends solely on the magnon-phonon interaction. (b) Taking into account finite interfacial thermal resistances for both spins and phonons results in a qualitatively different temperature profile.}
\label{fig:Tmp}%
\end{figure}

We will now turn to the temperature of the magnetic excitations (the magnons) in the ferromagnet, generalizing the work of Sanders and Walton~\cite{Sanders1977} who applied a model initially proposed by Kaganov \textit{et al.}~\cite{Kaganov1957} to a coupled system of magnons and phonons.\\
Let $\Delta T_\mathrm{mp}$ denote the difference between the magnon temperature $T_\mathrm{m}$ and the phonon temperature $T_\mathrm{p}$, then the magnon-phonon relaxation time $\tau_\mathrm{mp}$ is defined as
\begin{equation}
	\dfrac{}{t}\Delta T_\mathrm{mp}=-\frac{\Delta T_\mathrm{mp}}{\tau_\mathrm{mp}},
\label{eq:mprelax}
\end{equation}
and the time evolution of $T_\mathrm{m}$ and $T_\mathrm{p}$ follows
\begin{equation}
	\begin{split}
		\dfrac{T_\mathrm{p}}{t}&=\frac{c_\mathrm{m}}{c_\mathrm{t}}\frac{T_\mathrm{m}-T_\mathrm{p}}{\tau_\mathrm{mp}},\\
		\dfrac{T_\mathrm{m}}{t}&=\frac{c_\mathrm{p}}{c_\mathrm{t}}\frac{T_\mathrm{p}-T_\mathrm{m}}{\tau_\mathrm{mp}},
	\end{split}
\label{eq:mptime}
\end{equation}
where $c_\mathrm{m}$, $c_\mathrm{p}$ and $c_\mathrm{t}=c_\mathrm{p}+c_\mathrm{m}$ denote the magnon, phonon and total (sum of the two) heat capacity per unit volume. The heat flux $Q_\mathrm{mp}$ between the phonon and the magnon system is then given by
\begin{equation}
	\begin{split}
		Q_\mathrm{mp}(x)&=c_\mathrm{m}\dfrac{T_\mathrm{m}(x)}{t}\\
		&=\frac{c_\mathrm{p}c_\mathrm{m}}{c_\mathrm{t}}\frac{T_\mathrm{p}(x)-T_\mathrm{m}(x)}{\tau_\mathrm{mp}},
	\end{split}
\label{eq:ptomheatflow}
\end{equation}
where $x$ is the position along the thermal gradient (\textit{cf.} Fig.~\ref{fig:thSSE} and Fig.~\ref{fig:Tmp}). According to Eq.~\eqref{eq:heatSS}, the magnon temperature obeys
\begin{equation}
	 \dfrac{^2T_\mathrm{m}(x)}{x^2}+\frac{c_\mathrm{p}c_\mathrm{m}}{c_\mathrm{t}}\frac{1}{\kappa_\mathrm{m}\tau_\mathrm{mp}}\left[T_\mathrm{p}(x)-T_\mathrm{m}(x)\right]=0,
\label{eq:Tm}
\end{equation}
where $\kappa_\mathrm{m}$ is the magnon thermal conductivity. The phonon temperature is given by
\begin{equation}
	 \dfrac{^2T_\mathrm{p}(x)}{x^2}+\frac{c_\mathrm{p}c_\mathrm{m}}{c_\mathrm{t}}\frac{1}{\kappa_\mathrm{p}\tau_\mathrm{mp}}\left[T_\mathrm{m}(x)-T_\mathrm{p}(x)\right]=0.
\label{eq:Tp}
\end{equation}
We now consider an insulating ferromagnet (YIG) of length $d$ enclosed to the left $L$ ($x=0$) and right $R$ ($x=d$) by two nonmagnetic materials (i.e., the substrate on the left and the normal metal on the right as shown in Fig.~\ref{fig:Tmp}). We furthermore assume fixed phonon temperatures $T_\mathrm{L}^\mathrm{p}=\text{const.}$ for $x=0$ and $T_\mathrm{R}^\mathrm{p}=\text{const.}$ for $x= d$. When the material on the left or right side of the insulating ferromagnet is metallic, we take $T_\mathrm{L/R}=T_\mathrm{L/R}^\mathrm{p}=T_\mathrm{L/R}^\mathrm{e}$ for the sake of simplicity. As discussed at the end of Sec.~\ref{sec:phonheat} this constraint will not be utilized for the 3D FEM simulations in Secs.~\ref{sec:thermlandsc} and~\ref{sec:transSSE}. According to Eq.~\eqref{eq:thermbound} the boundary conditions for the phonon temperature are
\begin{equation}
	\begin{split}
		-\kappa_\mathrm{p}\left.\dfrac{T_\mathrm{p}(x)}{x}\right|_{x=0}&=\frac{1}{R_\mathrm{th,L}^\mathrm{p}}\left[T_\mathrm{L}-T_\mathrm{p}(0)\right],\\
		-\kappa_\mathrm{p}\left.\dfrac{T_\mathrm{p}(x)}{x}\right|_{x=d}&=\frac{1}{R_\mathrm{th,R}^\mathrm{p}}\left[T_\mathrm{p}(d)-T_\mathrm{R}\right],
	\end{split}
\label{eq:pbound}
\end{equation}
with the appropriate interfacial thermal conductances $(R_\mathrm{th,L}^\mathrm{p})^{-1}$ and $(R_\mathrm{th,R}^\mathrm{p})^{-1}$ for the left and right interface (\textit{cf.} App.~\ref{sec:Kapitza}). Magnons cannot exist in nonmagnetic materials. In spin pumping~\cite{Tserkovnyak2002a} and spin Seebeck experiments, however, the spin current, i.e., the transfer of angular momentum across the ferromagnet/normal metal interface, is accompanied by an energy transfer~\cite{Brataas2008} and thus an interface magnetic heat current $q_\mathrm{me}=\left(R_\mathrm{th}^\mathrm{m}\right)^{-1} \dT$ proportional to the interface magnetic heat conductance:~\cite{Xiao2010,pcTserkovnyak}
\begin{equation}
	\left(R_\mathrm{th}^\mathrm{m}\right)^{-1}=\frac{k_\mathrm{B}T}{\hbar}\frac{\mu_\mathrm{B}k_\mathrm{B}\gsm\eta}{\pi M_\mathrm{s}V_\mathrm{a}},
\label{eq:Gm}
\end{equation}
where $\mu_\mathrm{B}$ is the Bohr magneton. In the macrospin approximation the characteristic energy in Eq.~\eqref{eq:Gm} is given by the ferromagnetic resonance (FMR) frequency of the spin system,~\cite{Xiao2010} however, in a system of thermal magnons this should be the temperature of the magnons.~\cite{pcTserkovnyak} The boundary conditions for the magnon system are:
\begin{equation}
	\begin{split}
		-\kappa_\mathrm{m}\left.\dfrac{T_\mathrm{m}(x)}{x}\right|_{x=0}&=\frac{1}{R_\mathrm{th,L}^\mathrm{m}}\left[T_\mathrm{L}-T_\mathrm{m}(0)\right],\\
		-\kappa_\mathrm{m}\left.\dfrac{T_\mathrm{m}(x)}{x}\right|_{x=d}&=\frac{1}{R_\mathrm{th,R}^\mathrm{m}}\left[T_\mathrm{m}(d)-T_\mathrm{R}\right].
	\end{split}
\label{eq:mbound}
\end{equation}
With these boundary conditions, $T_\mathrm{m}(x)$ and $T_\mathrm{p}(x)$ can be calculated from Eqs.~\eqref{eq:Tm} and~\eqref{eq:Tp}. Note that a similar system of equations was solved in Ref.~\onlinecite{Xiao2010} with identical interfaces L and R. The present approach enables the description of a large number of experiments with very different boundary conditions for the substrate/ferromagnet and ferromagnet/normal metal interface.\\
Fig.~\ref{fig:Tmp} sketches $T_\mathrm{m}(x)$ and $T_\mathrm{p}(x)$ profiles as obtained from Eqs.~\eqref{eq:Tm} and~\eqref{eq:Tp} in different limits. When the phonon interfacial thermal resistance and interface magnetic heat current are disregarded, we recover the result of Ref.~\onlinecite{Sanders1977} in which $\dT$ is exclusively governed by the magnon-phonon interaction [Fig.~\ref{fig:Tmp}(a)]. Taking into account the phonon interfacial thermal resistance and the interface magnetic heat conductance, qualitatively different temperature profiles emerge [Fig.~\ref{fig:Tmp}(b)].

To calculate the temperature profiles for the coupled phonon-electron systems in the metallic layer Eqs.~\eqref{eq:Tm}--\eqref{eq:pbound} and~\eqref{eq:mbound} can simply be modified by substituting the magnon parameters ($T_\mathrm{m},\ c_\mathrm{m},\ \kappa_\mathrm{m},\ \tau_\mathrm{mp},\ R_\mathrm{th,L/R}^\mathrm{m}$) with the appropriate electron ones ($T_\mathrm{e},\ c_\mathrm{e},\ \kappa_\mathrm{e},\ \tau_\mathrm{ep},\ R_\mathrm{th,L/R}^\mathrm{e}$).

In the following sections, we address the phonon and magnon temperatures in YIG films exposed to a thermal gradient, and then consider the $T_\mathrm{p},\ T_\mathrm{e}$ and $T_\mathrm{m}$ profiles under local heating.

\section{One-dimensional temperature profiles}\label{sec:numeric}
\begin{figure}%
\includegraphics[width=\columnwidth]{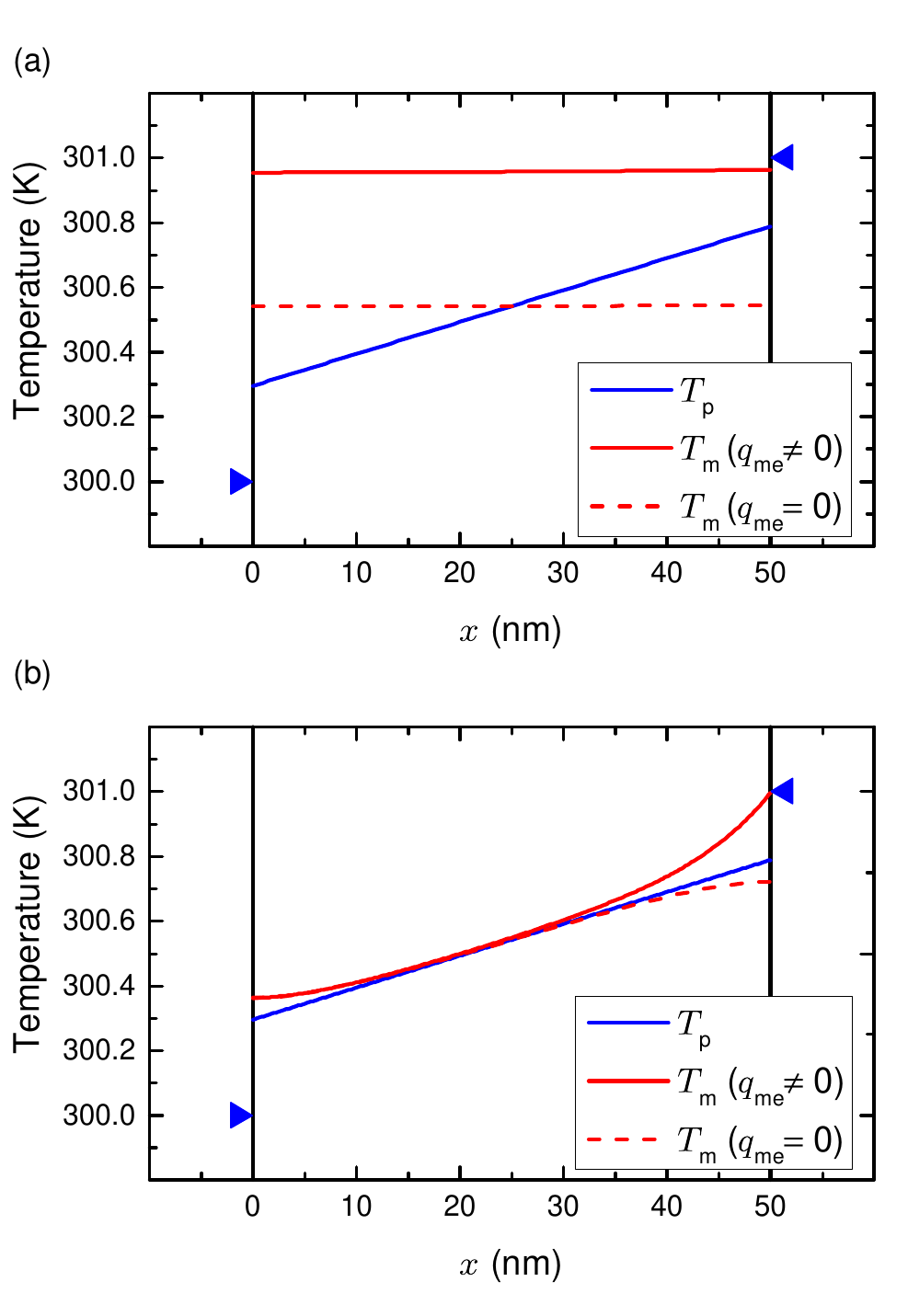}%
\caption{(Color online) (a) Magnon and phonon temperature profiles in the macrospin approximation at microwave frequencies ($\omega\approx\SI{20}{\giga Hz}$, \textit{cf.} Ref.~\onlinecite{Xiao2010}) calculated from Eq.~\eqref{eq:Tm} and Eq.~\eqref{eq:Tp} for a $\SI{50}{\nano m}$ thick YIG film with $T_\mathrm{L}=\SI{300}{K}$ and $T_\mathrm{R}=\SI{301}{K}$, the material parameters from Tab.~\ref{tab:matpar} and appropriate $(R_\mathrm{th})^{-1}$ for both phonons and magnons. The interface magnetic heat current $q_\mathrm{me}$ is limited to the right interface here. The dashed line depicts the case when the interface magnetic heat current $q_\mathrm{me}$ is not taken into account. (b) Same as (a) but allowing for thermal excitation of magnons with arbitrary wave lengths. While the magnon temperature profiles appear to be qualitatively different and $\dT$ is smaller right at the interface, all magnons within a finite length ($\sqrt[3]{V_\mathrm{a}}\approx\SI{1.3}{\nano m}$), contribute to the pumped spin current such that the effective $\dT$ is only moderately reduced.}%
\label{fig:MagTemp}%
\end{figure}

In order to quantitatively calculate $T_{\mathrm{p}}(x)$ and $T_{\mathrm{m}}(x)$ in YIG thin films from Eqs.~\eqref{eq:Tm}--\eqref{eq:mbound} the magnon parameters $c_\mathrm{m},\ \kappa_\mathrm{m}$, and $\tau_\mathrm{mp}$ in YIG are required, but to the best of our knowledge are only well established for temperatures $T\lesssim\SI{10}{K}$.\\
The available low temperature data~\cite{Bhandari1966,Pan2013} for the YIG magnon thermal conductivity show that the magnonic contribution to the total thermal conductivity~\cite{Sanders1977} is of the order of a few percent at low temperatures. However, with the notable exception of spin ladder and spin chain systems,~\cite{ElHaes2004,Montagnese2013} it is generally assumed that the magnonic contribution to the total thermal conductivity at room temperature~\cite{Xiao2010,Douglass1963} is very small. Theory~\cite{Akhiezer1968,Landau1981} indeed predicts $\kappa_\mathrm{m}$ to decay inversely proportional to $T$ or even exponentially at elevated temperatures due to increasing scattering processes.~\cite{Kolokolov1984} Additional support for very small $\kappa_\mathrm{m}$ in YIG comes from an analysis~\cite{Padture1997} of the total thermal conductivity that does not show any significant features around the Curie temperature where the relative change in the magnon thermal conductivity should be large. Due to the aforementioned reasons and for lack of better data we here assume $\kappa_\mathrm{m}$ to be of the order of $10^{-2}-\SI{e-3}{W/(m K)}$, which is also supported by earlier theoretical estimates for $\kappa_\mathrm{m}$ at elevated temperatures,~\cite{Douglass1963} and we will use the mean of the assumed range of $\kappa_\mathrm{m}=\SI{3e-3}{W/(m K)}$ for our calculations.
The expression for $\kappa_\mathrm{m}$ adopted in Ref.~\onlinecite{Xiao2010} is limited to the low temperature regime and yields values at room temperature of $\kappa_\mathrm{m}>\SI{1e4}{W/(m\,K)}$ which appear odd based on the available data.\\
We calculate the magnon heat capacity from the spin wave stiffness $D=\SI{8.5e-40}{J\,m^2}$ (Ref.~\onlinecite{Cherepanov1993,Srivastava1987}):~\cite{Sato1955,Xiao2010}
\begin{equation}
	c_\mathrm{m}=\frac{15\zeta(5/2)}{32}\sqrt{\frac{k_\mathrm{B}^5T^3}{\pi^3D^3}}
\label{eq:CmAFi}
\end{equation}
and obtain a value of $c_\mathrm{m}\approx\SI{16750}{J/(m^3K)}$ at $T=\SI{300}{K}$.\\
The magnon-phonon relaxation time $\tau_\mathrm{mp}$ critically depends on the specific magnon mode. While it is relatively large for microwave magnons~\cite{Spencer1962,Dzyapko2010} it decreases significantly for short wavelength, thermal magnons.~\cite{Kumar1967} Assuming that the majority of the magnetic damping in the YIG is due to the interaction with phonons, one can estimate $\tau_\mathrm{mp}$ by (\textit{cf.} App.\ref{sec:tmp})
\begin{equation}
	\tau_\mathrm{mp}\approx\frac{\hbar}{\alpha_\mathrm{G}k_\mathrm{B}T},
\label{eq:tmp}
\end{equation}
where $\alpha_\mathrm{G}$ is the Gilbert damping parameter of the bare YIG film. As in Eq.~\eqref{eq:Gm}, the expression for $\tau_\mathrm{mp}$ differs for the macrospin-approximation (\textit{cf.} Ref.~\onlinecite{Xiao2010}) and for a magnon system, however, the above should be more appropriate in the case of thermal magnons in an extended ferromagnet.  While there is a large spread in of reported values for $\alpha_\mathrm{G}\approx10^{-3}-10^{-5}$ (Ref.~\onlinecite{Althammer2012, Heinrich2011, Hung2010, Kurebayashi2011, Kajiwara2010, Hoffman2013}) higher values are generally found in thin films where two and three magnon scattering processes contribute to the damping. We therefore adopt $\alpha_\mathrm{G}=10^{-4}$ in the following and obtain a magnon-phonon relaxation time for thermal magnons of $\tau_\mathrm{mp}=\SI{255}{\pico s}$.\\
\newcolumntype{d}{D{.}{.}{1.3}}
\newcolumntype{k}{D{.}{.}{2.2}}
\begin{table*}
\begin{ruledtabular}
\begin{tabular}{cccccccc}
 & mass & phonon heat & electron heat & phonon thermal & electron thermal & longitudinal & transverse\\
 & density & capacity & capacity & conductivity & conductivity & speed of sound & speed of sound\\
 & $\varrho\ (\SI{}{kg m^{-3}})$ & $C_\mathrm{p}\ (\SI{}{J \kilo g^{-1} K^{-1}})$ & $C_\mathrm{e}\ (\SI{}{J \kilo g^{-1} K^{-1}})$ & $\kappa_\mathrm{p}\ (\SI{}{W m^{-1} K^{-1}})$ & $\kappa_\mathrm{e}\ (\SI{}{W m^{-1} K^{-1}})$ & $v_\mathrm{long}\ (\SI{}{m s^{-1}})$ & $v_\mathrm{trans}\ (\SI{}{m s^{-1}})$\\
\hline
Pt & $21450$~\textsuperscript{a} & $120$~\textsuperscript{a,e} & $10$~\textsuperscript{e} & $8$~\textsuperscript{a,f} & $64$~\textsuperscript{a,f} & $3300$~\textsuperscript{a} & $1700$~\textsuperscript{a}\\
YIG & $5170$~\textsuperscript{b} & $570$~\textsuperscript{c} & - & $6$~\textsuperscript{g} & - & $7170$~\textsuperscript{b} & $3843$~\textsuperscript{b}\\
GGG & $7080$~\textsuperscript{c} & $400$~\textsuperscript{c} & - & $8$~\textsuperscript{h} & - & $6545$~\textsuperscript{k} & $3531$~\textsuperscript{k}\\
YAG & $4550$~\textsuperscript{d} & $625$~\textsuperscript{c} & - & $9$~\textsuperscript{g} & - & $8600$~\textsuperscript{l} & $4960$~\textsuperscript{l}\\
Au & $19300$~\textsuperscript{a} & $129$~\textsuperscript{a,e} & $1$~\textsuperscript{e} & $2$~\textsuperscript{i,j} & $316$~\textsuperscript{i,j} & $3240$~\textsuperscript{a} & $1200$~\textsuperscript{a}
\end{tabular}
\end{ruledtabular}
\footnotesize{\textsuperscript{a} Ref.~\onlinecite{Lide2008}\hspace{2cm}\textsuperscript{b} Ref.~\onlinecite{Clark1961}\hspace{2cm}\textsuperscript{c} Ref.~\onlinecite{Hofmeister2006}\hspace{2cm}\textsuperscript{d} Ref.~\onlinecite{Ikesue1995}\hspace{2cm}\textsuperscript{e} Ref.~\onlinecite{Lin2008}\hspace{2cm}\textsuperscript{f} Ref.~\onlinecite{Duggin1970}\hspace{2cm}\textsuperscript{g} Ref.~\onlinecite{Padture1997}\hspace{2cm}\textsuperscript{h} Ref.~\onlinecite{Sirota1992}\hspace{2cm}\textsuperscript{i} Ref.~\onlinecite{Fiege1999}\hspace{2cm}\textsuperscript{j} Ref.~\onlinecite{Goicochea2011}\hspace{2cm}\textsuperscript{k} Ref.~\onlinecite{Zhou2011}\hspace{2cm}\textsuperscript{l} Ref.~\onlinecite{Kitaeva1985}}
\caption{
Material parameters used for the calculation of the phonon temperature distribution in YIG/Pt-type hybrids. Electronic contributions to the values for the phonon heat capacity and the phonon thermal conductivity in platinum and gold have been separated using the quoted sources. Additionally $C_\mathrm{p}\gg C_\mathrm{m}$ and $\kappa_\mathrm{p}\gg \kappa_\mathrm{m}$ (Ref.~\onlinecite{Xiao2010}) so that heat capacity and thermal conductivity in the YIG can be considered essentially free from magnonic contributions.}
\label{tab:matpar}
\end{table*}
The different temperature profiles obtained from the macrospin approximation and for a magnon system are displayed in in Fig.~\ref{fig:MagTemp} for a $\SI{50}{\nano m}$ thick YIG film with $T_\mathrm{L}=\SI{300}{K}$ and $T_\mathrm{R}=\SI{301}{K}$ using the material parameters listed in Tab.~\ref{tab:matpar}. Figure~\ref{fig:MagTemp}(a) depicts $T_{\mathrm{m}}$ and $T_{\mathrm{p}}$ calculated from Eqs.~\eqref{eq:Tm}~\eqref{eq:Tp} in the macrospin model at microwave frequencies [$\omega\approx\SI{20}{\giga Hz},\ \tau_\mathrm{mp}\approx\SI{0.4}{\micro s}$ (Ref.~\onlinecite{Spencer1962})]. Here the magnon temperature is essentially constant over the length of the thin film. While the interface magnetic heat current $q_\mathrm{me}$ across the ferromagnet/normal metal interface is still relatively small at microwave frequencies $\dT$ is already notably reduced to about $\SI{37}{\milli K}$. Taking $\left(R_\mathrm{th}^\mathrm{m}\right)^{-1}$ and $\tau_\mathrm{mp}$ for thermal magnons yields the profiles depicted in Fig.~\ref{fig:MagTemp}(b). Due to the significantly stronger interaction between magnons and phonons ($\propto\tau_\mathrm{mp}^{-1}$) the magnon temperature approaches the phonon temperature even over very short lengthscales and also the interface magnetic heat current is much stronger here. However, in the formalism by Xiao \textit{et al.},\cite{Xiao2010} the magnetic coherence length $\sqrt[3]{V_\mathrm{a}}\approx\SI{1.3}{\nano m}$ gives the length over which a given perturbation is felt, or in other words, the effective width of the interface and hence the length over which magnons contribute to the pumped spin current. This results in $\dT$ being reduced from $37$ to about $\SI{27}{\milli K}$.  Hence, although the magnon temperature profile drastically changes from Fig.~\ref{fig:MagTemp}(a) to~\ref{fig:MagTemp}(b), the effect on the effective $\dT$ at the interface is rather weak.

The effect of the interface magnetic heat current $q_\mathrm{me}$ on $\dT$ scales inversely proportional with the thickness of the ferromagnetic layer up to a critical thickness. For thermal magnons, however, this scaling is limited to few $\SI{}{\nano m}$, compared to about $\SI{100}{\nano m}$ in the macrospin limit.

The magnetic coherence length $\sqrt[3]{V_\mathrm{a}}$ is closely related with the thermal de Broglie length of the magnon system. In contrast to the bulk magnon model considered by Xiao \textit{et al.},~\cite{Xiao2010} Hoffman \textit{et al.}~\cite{Hoffman2013} presented an approach based on a minimal Landau-Lifshitz-Gilbert treatment of films with arbitrary thickness. This approach proceeds from the assumption that the magnon-phonon relaxation is described by the Gilbert damping, but does not take into account the Kapitza resistance. For sufficiently thick magnetic layers Hoffman \textit{et al.} find a spin current $j_\mathrm{s}$ across the ferromagnet/normal metal interface of
\begin{equation}
	j_\mathrm{s}=\frac{\hbar\gsm\gamma}{4\pi M_\mathrm{s}}\frac{\left(\frac{k_\mathrm{B}T}{D}-\frac{\gamma\hbar\mu_0 H}{D}\right)^{3/2}}{3\pi^2\left(1+\frac{2\hbar\gsm\gamma}{4\pi M_\mathrm{s}\alpha_\mathrm{G}d}\right)}k_\mathrm{B}\Delta T
\label{eq:Hoff}
\end{equation}
where $\Delta T$ is the (phonon) temperature drop across the ferromagnetic layer of thickness $d$ and $\mu_0 H$ with the vacuum permeability $\mu_0=4\pi\times\SI{e-7}{V s/(A m)}$ is the externally applied magnetic field ($\SI{70}{\milli T}$ in our case). Substituting the first term in Eq.~\eqref{eq:SSE} with Eq.~\eqref{eq:Hoff} and using identical parameters, the theory by Hoffman \textit{et al.} agree within an order of magnitude. For the case discussed above, the theory by Hoffman \textit{et al.} yields smaller values, but our calculations show that the magnon temperature gradient at the ferromagnet/normal metal interface is not equal to the phonon one.

In summary of this section, we found that in thin films the interface magnetic heat conductance can have a substantial impact on $T_\mathrm{m}$ and that for both the macrospin model and thermal magnons a similar temperature difference at the ferromagnet/normal metal interface arises. A comparison of the theories by Hoffman \textit{et al.}~\cite{Hoffman2013} and Xiao \textit{et al.}~\cite{Xiao2010} shows reasonable agreement and underlines the importance of the magnon-phonon interaction. 
We would like to stress that within a reasonable range (at least of one order of magnitude for each quantity) of possible values for $c_\mathrm{m},\ \kappa_\mathrm{m}$ and $\tau_\mathrm{mp}$ the results presented in this section do not change qualitatively. As pointed out above, however, the chosen values for $c_\mathrm{m},\ \kappa_\mathrm{m}$ and $\tau_\mathrm{mp}$ have not been experimentally determined at room temperature, but were estimated from theoretical calculations. Also, the fact that the magnon temperature profiles vary notably over the magnetic coherence length could indicate the limits of the diffusive Sanders and Walton~\cite{Sanders1977} approach.

\section{Temperature profiles in three-dimensional thermal landscapes}\label{sec:thermlandsc}
\begin{figure}%
\includegraphics[width=\columnwidth]{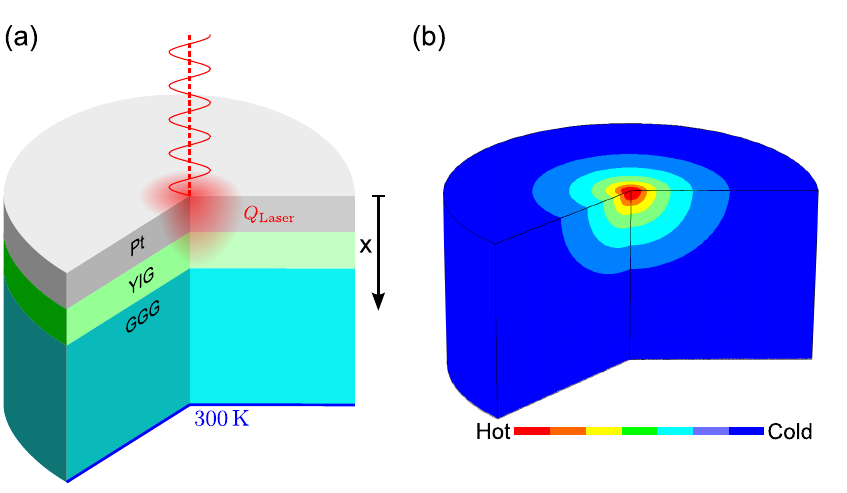}%
\caption{(Color online) (a) Depiction of the initial setup of the problem in the 3D FEM (not to scale). The bottom of the substrate is fixed at $T=\SI{300}{K}$, whereas the other outer borders are thermally insulating. At the Pt/YIG and YIG/GGG interfaces the interfacial thermal resistances calculated in Sec.~\ref{sec:Kapitza} are applied. (b) Cut through a typical result obtained from a steady state simulation of the heat transfer problem with a logarithmic and capped scale of the phonon temperature rise in the simulated sample.}%
\label{AnsysSetup}%
\end{figure}

For the description of our local spin Seebeck experiments described in more detail in Refs.~\onlinecite{Weiler2012,Schreier2012}, in which a focused laser beam is used to locally heat YIG/Pt hybrid samples, a 1D temperature model is not sufficient. We therefore use 3D FEM to simulate the temperature distribution in the experiments. In the 3D FEM, the geometry~\footnote{We employ the cylindrical symmetry of the problem to speed up the calculations.} of the problem is set up first (Fig.~\ref{AnsysSetup}).
The 3D FEM allows us to couple the heat equations for the phonons, electrons, and magnons as given by Eqs.~\eqref{eq:Tm} and~\eqref{eq:Tp} and calculate the temperature profiles for three systems simulatenously. The heating by the laser light, which we here assume to be exclusively absorbed by the electrons in the metal, is given by

\begin{figure}%
\includegraphics[width=\columnwidth]{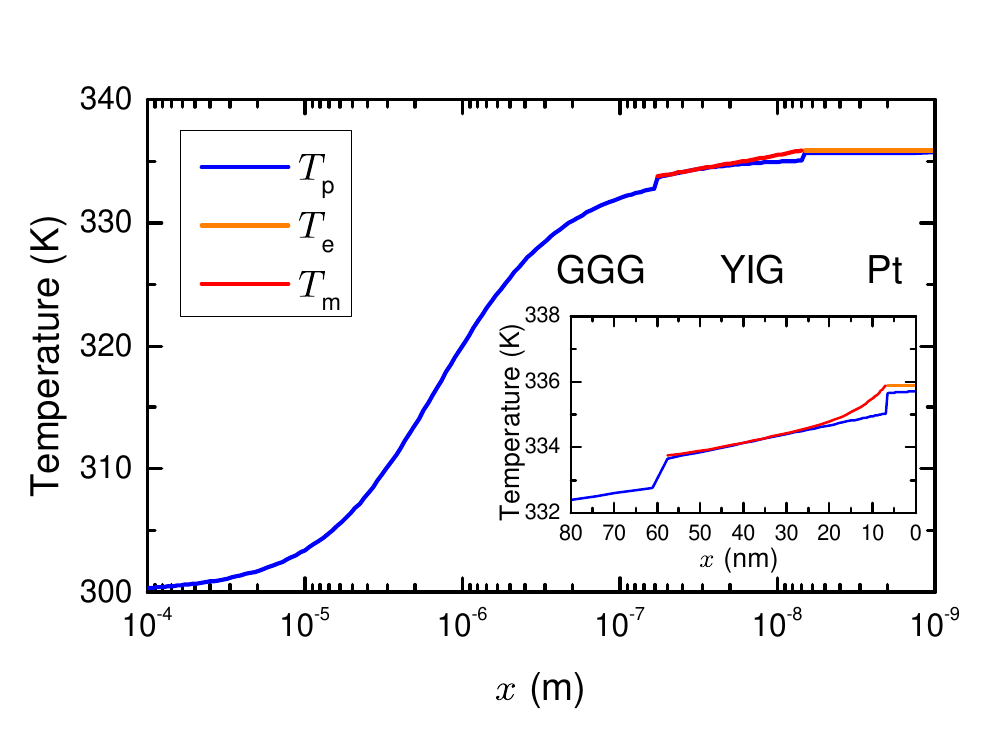}%
\caption{(Color online) Phonon, electron and magnon temperature distributions along the optical axis of the laser beam at $\SI{10}{\milli W}$ laser power in a GGG($\SI{500}{\micro m}$)/YIG($\SI{54}{\nano m}$)/Pt($\SI{7}{\nano m}$) sample calculated via 3D FEM for the entire sample. The inset shows the thin film region.}%
\label{fig:ANSRG}%
\end{figure}

\begin{widetext}
\begin{eqnarray}
		 P_{\mathrm{inc},i}(x,r)&=&P_\mathrm{Laser}\left[\prod_{j=1}^i{(1-R_j)}\right]\exp\left[-\alpha_i\left(x-\sum_{j=1}^{i-1}{t_j}\right)-\sum_{j=1}^{i-1}{\alpha_jt_j}\right]\exp\left(-2\frac{r^2}{a^2}\right),\\
		P_{\mathrm{ref},i}(x,r)&=&P_{\mathrm{inc},i}(t_i,r)R_{i+1}\exp\left[\alpha_i\left(x_i-\sum_{j=1}^i{t_j}\right)\right],\\
		Q_i(x,r)&=&\frac{2}{\pi a^2}\left[-\pfrac{P_{\mathrm{inc},i}(x,r)}{x}+\pfrac{P_{\mathrm{ref },i}(x,r)}{x}\right]\label{eq:PwrInLayer},
\end{eqnarray}
\end{widetext}
which is a modified version of the expression found in Ref.~\onlinecite{Reichling1994} that accounts for both the incident ($P_{\mathrm{inc}}$) and reflected ($P_{\mathrm{ref}}$) parts of the laser light.
Here $P_\mathrm{Laser},\ R_i,\ \alpha_i,\ t_i,\ a,\ x$ and $r$ denote the initial (optical) laser power, the reflectivity of the individual surfaces, the optical absorption coefficients (\textit{cf.} Tab.~\ref{tab:optpar}), the layer thicknesses, the laser spot radius and the two coordinates of cylindrical symmetry,~\footnote{We deviate from the standard convention $(z,r)$ for cylindrical coordinates for consistency in this paper.} respectively. The reflectivity $R_i$ at the interface of the layers $i-1$ and $i$ is computed using the Fresnel equation for normal incidence~\cite{Fresnel1815,Zinth1998}
\begin{equation}
	R_i=\left|\frac{n_{i-1}-n_i}{n_{i-1}+n_i}\right|^2,
\label{eq:FResnel}
\end{equation}
where $n_i$ denotes the complex refractive index (cf. Table~\ref{tab:optpar}) in layer $i$. The laser spot radius in our experiments is $a=\SI{2.5}{\micro m}$. However, our experimental results and geometric considerations show that the measured voltage signals do not depend on $a$ as long as the laser spot is located fully within the Hall bar.\\
\begin{table}
\begin{ruledtabular}
\begin{tabular}{ck@{$\times$\hspace{-1.2cm}}ld@{$+$\hspace{-1.6cm}}d}
& \multicolumn{2}{c}{\hspace{-.6cm}absorption} & \multicolumn{2}{c}{\hspace{-.6cm}refractive}\\
 & \multicolumn{2}{c}{\hspace{-.6cm}coefficient} & \multicolumn{2}{c}{\hspace{-.6cm}index}\\
 & \multicolumn{2}{c}{\hspace{-.6cm}$\alpha\ (\SI{}{m^{-1}})$} & \multicolumn{2}{c}{\hspace{-.6cm}$n$}\\
\hline
Pt & 82	&	$10^6$~\textsuperscript{a} & 2.41	&	4.3\mathrm{i}~\textsuperscript{a}\\
YIG & 0.5	&	$10^5$~\textsuperscript{b} & 2.27	&	0.003\mathrm{i}~\textsuperscript{b}\\
GGG & \approx0	&	10~\textsuperscript{c} & 1.96	&	0.0\mathrm{i}~\textsuperscript{d}\\
YAG & \approx0	&	10~\textsuperscript{a} & 1.83	&	0.0\mathrm{i}~\textsuperscript{a}\\
Au & 62.5	&	$10^6$~\textsuperscript{a} & 0.16	&	3.28\mathrm{i}~\textsuperscript{a}
\end{tabular}
\end{ruledtabular}
\footnotesize{\textsuperscript{a}~Reference~\onlinecite{Palik1985}\hspace{.3cm}\textsuperscript{b}~Reference~\onlinecite{Scott1974}\hspace{.3cm}\textsuperscript{c}~Reference~\onlinecite{Potera2010}\hspace{.3cm}\textsuperscript{d}~Reference~\onlinecite{Wood1990}}
\caption{
Optical material parameters at $\lambda_\mathrm{Laser}=\SI{660}{\nano m}$. The small absorption coefficient of YIG has been neglected in the simulations.}
\label{tab:optpar}
\end{table}
As an additional boundary condition, the lower end of the substrate is set to a fixed temperature of $\SI{300}{K}$ to simulate the effect of the copper heat sink the samples are attached to in experiment.~\cite{Weiler2012} For the coupling between the electrons in the platinum and gold layer with the phonons we use an electron-phonon relaxation time of $\tau_\mathrm{ep}^\mathrm{Pt}=\SI{1.8}{\pico s}$ and $\tau_\mathrm{ep}^\mathrm{Au}=\SI{1.9}{\pico s}$, respectively (both Ref.~\onlinecite{Lin2008}). Black body radiation and convective cooling are not taken into account as their contribution in this particular geometry is much smaller than the effect of heat flow within the sample, as evident from the following estimations based on the Stefan-Boltzmann law and Newton's law of cooling:~\cite{Jiji2009}
\begin{eqnarray}
	P_\mathrm{rad}&=&\sigma A (T_\mathrm{sample}^4-T_\mathrm{env}^4),\\
	P_\mathrm{conv}&=&h A (T_\mathrm{sample}-T_\mathrm{env}).
\end{eqnarray}
Using the Stefan-Boltzmann constant $\sigma=5.67\times10^{-8}\SI{}{W/(m^2\,K^4)}$, the Hall bar surface $A=80\times\SI{950}{\micro m^2}$, the heat transfer coefficient for air $h\leq\SI{30}{W/(m^2 K)}$~\cite{Jiji2009} and $T_\mathrm{sample}\lessapprox\SI{400}{K}$ we find that less than $\SI{0.1}{\milli W}$ are lost due to radiation and convective cooling which is less than $1\%$ of the power absorbed by the sample for typical experimental values (\textit{cf.} Sec.~\ref{sec:experiment}). The 3D FEM then yields the phonon, electron, and magnon temperature distributions also for local laser heating of YIG/Pt-type hybrids based on the procedure outlined in Sec.~\ref{sec:phonheat}. 
Figure~\ref{fig:ANSRG} shows the phonon, electron, and magnon temperature profiles at the center of the laser spot along the film normal. As in the one-dimensional case $\dT$ is averaged over the magnetic coherence length. The inclusion of interfacial thermal resistance and the separate treatment of electrons and phonons in the platinum lead to a substantial increase in $\dT$ by about an order of magnitude.\\
Since only a small area is heated in our experiments a single $\dT$ can not be given, due to the lateral interaction of the individual systems. Additionally, the current that is eventually created by the spin Seebeck and inverse spin Hall effect is effectively short-circuited by the non heated region such that one needs to substitute the term $l\dT$ in Eq.~\eqref{eq:SSE} by the integral expression
\begin{equation}
	l\dT=\frac{2\pi}{w}\int\dT(r)rdr,
\label{eq:ldT}
\end{equation}
where $w=\SI{80}{\micro m}$ is the width of the Hall bar.
Using this $l\dT$ one can now compare the spin Seebeck effect expected from theory with experiment. In the following comparison we will use $l\dT=|l\dT|$ for clarity.
\begin{figure}%
\includegraphics[width=\columnwidth]{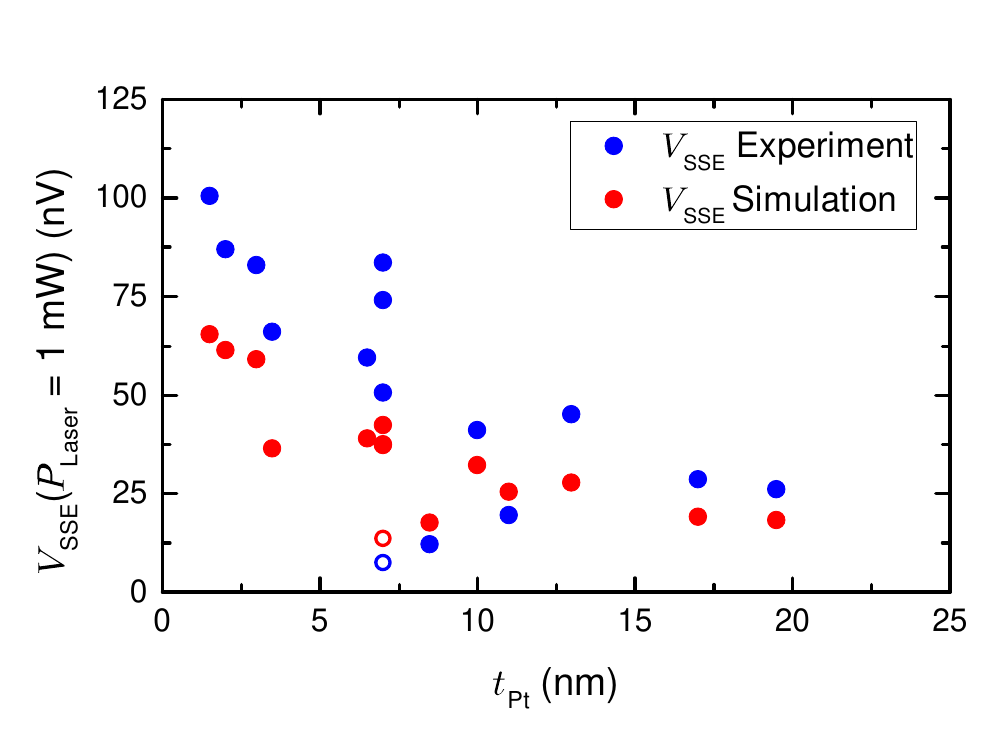}%
\caption{(Color online) Comparison of the observed and computed spin Seebeck voltages under local laser heating. The $\dT$ in Eq.~\eqref{eq:SSE} is calculated with the sample parameters from Tabs.~\ref{tab:matpar}--\ref{tab:samppar} and the magnon properties outlined in Sec.~\ref{sec:numeric}. The open circles depict the sample with an additional gold layer between the platinum and the YIG. Generally good agreement between theory and experiment is found.}%
\label{fig:theoexp}%
\end{figure}
\newcolumntype{p}{D{.}{.}{3.1}}
\begin{table}
\begin{ruledtabular}
\begin{tabular}{lpd}
	Sample & \multicolumn{1}{c}{$\rho (\SI{}{\nano\ohm m})$} & \multicolumn{1}{c}{$l\dT(\SI{e-9}{K m})$}\\
	\hline
	GGG/YIG(50)/Pt(7) & 409.4 & 1.35\\
	GGG/YIG(54)/Pt(7) & 406.5 & 1.35\\
	GGG/YIG(46)/Pt(3.5) & 306.6 & 0.96\\
	GGG/YIG(58)/Pt(2) & 761.7 & 0.78\\
	GGG/YIG(56.5)/Pt(1.5) & 1089.9 & 0.73\\
	GGG/YIG(61)/Pt(11) & 334.5 & 1.63\\
	GGG/YIG(53)/Pt(8.5) & 348.3 & 0.85\\
	GGG/YIG(52)/Pt(17) & 331.7 & 1.90\\
	YAG/YIG(59)/Pt(7) & 487.7 & 1.36\\
	YAG/YIG(64)/Pt(3) & 622.2 & 0.92\\
	YAG/YIG(61)/Pt(19.5) & 358.7 & 1.97\\
	YAG/YIG(63)/Pt(6.5) & 412.0 & 1.31\\
	YAG/YIG(60)/Pt(10) & 429.0 & 1.58\\
	YAG/YIG(60)/Pt(13) & 434.9 & 1.75\\
	GGG/YIG(15)/Au(7)/Pt(7) & 143.0 & 2.19\\
\end{tabular}
\end{ruledtabular}
\caption{Samples used in this study. Numbers in parentheses indicate layer thickness in nm (rounded to the next $\SI{5}{\angstrom}$). $l \dT$ denotes the integrated temperature difference between the magnons and the electrons at the YIG/Pt (YIG/Au) interface at $P_\mathrm{Laser}=\SI{1}{\milli W}.$}
\label{tab:samppar}
\end{table}
Figure~\ref{fig:theoexp} shows a comparison between the voltages measured in our local laser heating setup (cf. Sec.~\ref{sec:experiment}) and theoretical values obtained from Eq.~\eqref{eq:SSE} for the YIG/Pt heterostructures listed in Table~\ref{tab:samppar}. For a YIG($\SI{54}{\nano m}$)/Pt($\SI{7}{\nano m}$) sample (also shown in Fig.~\ref{fig:theoexp}), a voltage $V_\mathrm{SSE}=\SI{74}{\nano V}$ is observed for $P_\mathrm{laser}=\SI{1}{\milli W}$ at the sample surface. Using $\gsm=\SI{1e19}{m^{-2}}$,~\cite{Althammer2013,Chen2013a} $\theta_\mathrm{H}=0.11$,~\cite{Althammer2013,Chen2013a} $\lambda=\SI{1.5}{\nano m}$,~\cite{Althammer2013,Chen2013a} $\gamma=\SI{1.76e11}{Hz/T}$, $M_\mathrm{s}=\SI{140e3}{A/m}$,~\cite{Dorsey1993} $D=\SI{8.5e-40}{J\,m^2}$ (Ref.~\onlinecite{Cherepanov1993,Srivastava1987}) and the value $l\dT=\SI{1.55e-9}{K m}$ obtained for $\SI{1}{\milli W}$ optical laser power from our numerical calculations, we obtain $V_\mathrm{SSE}=\SI{37}{\nano V}$ from Eq.~\eqref{eq:SSE} in good agreement with the experiment. This agreement is not limited to this particular sample as can be seen in Fig.~\ref{fig:theoexp}. Good agreement between experiment and theory is also found, for a YIG($\SI{15}{\nano m}$)/Au($\SI{7}{\nano m}$)/Pt($\SI{7}{\nano m}$) sample with $\rho=\SI{1.43e-7}{\ohm\,m}$. For this sample $V_\mathrm{SSE}=\SI{8}{\nano V}$ is measured in experiment and $l\dT=\SI{1.95e-9}{K m}$ corresponding to $V_\mathrm{SSE}=\SI{13}{\nano V}$ is obtained from our simulation using a spin mixing conductance of $\gsm=\SI{5e18}{m^{-2}}$ (Ref.~\onlinecite{Heinrich2011,Burrowes2012}) for the gold/yttrium iron garnet interface. Note that the computed value does not take into account any decrease of the spin current at the additional gold/platinum interface in this particular sample.\\
Generally this approach seems to slightly underestimate $\dT$ which could, however, be remedied by a slightly different magnon parameter set. Overall though, the spin Seebeck effect theory accounts for the experimental values, especially considering the uncertainties in the determination of $\dT$ as discussed above.\\
The simulation also show that it is unlikely that the measured voltages stem from the anomalous Nernst effect.~\cite{Huang2012} Using
\begin{equation}
	V_\mathrm{ANE}=-N_\mathrm{Nernst}\mu_0 M_\mathrm{s}\frac{2\pi}{w}\int\pfrac{T_\mathrm{e}(x,r)}{x}rdr,
\label{eq:VANE}
\end{equation}
with the Nernst coefficient $N_\mathrm{Nernst}$, the simulation shows that a Nernst coefficient of about $N_\mathrm{Nernst}\approx\SI{1e-3}{V/(K T)}$ is needed to explain the measured voltage of about $\SI{100}{\nano V}$ in the YIG($\SI{56.5}{\nano m}$)/Pt($\SI{1.5}{\nano m}$) sample at $P_\mathrm{Laser}=\SI{1}{\milli W}$. To exaggerate the anomalous Nernst effect we here also assumed that the entire platinum layer is evenly magnetized identical to the YIG, i.e. $M_\mathrm{s}^\mathrm{Pt}=M_\mathrm{s}^\mathrm{YIG}=\SI{140e3}{A/m}$ and does not decay exponentially within the first few monolayers.~\cite{Meservey1980} A Nernst coefficient of $N_\mathrm{Nernst}=\SI{1e-3}{V/(K T)}$ is, however, orders of magnitude larger than for instance the Nernst coefficient in bulk nickel of about $N_\mathrm{Nernst}^\mathrm{Ni}=\SI{5e-7}{V/(K T)}$ (Ref.~\onlinecite{Smith1911}) and cannot be motivated for magnetized platinum. We therefore conclude that potential contributions from the anomalous Nernst effect do not play any significant role in our measurements as already found in Ref.~\onlinecite{Weiler2012}.\\
We also would like to point out that the inclusion of the interfacial thermal resistance, in principle, allows us to scale $\dT$ without changing the thermal gradient $\pfrac{T_e}{x}$ in the thin films. This means that one could imagine a set of samples with identical $\pfrac{T_e}{x}$ but different $\dT$ such that the spin Seebeck effect ($\propto\dT$) and anomalous Nernst effect ($\propto\pfrac{T_e}{x}$) can unambiguously be disentangled.\\

\section{Transverse temperature profiles}\label{sec:transSSE}
\begin{figure}%
\includegraphics[width=\columnwidth]{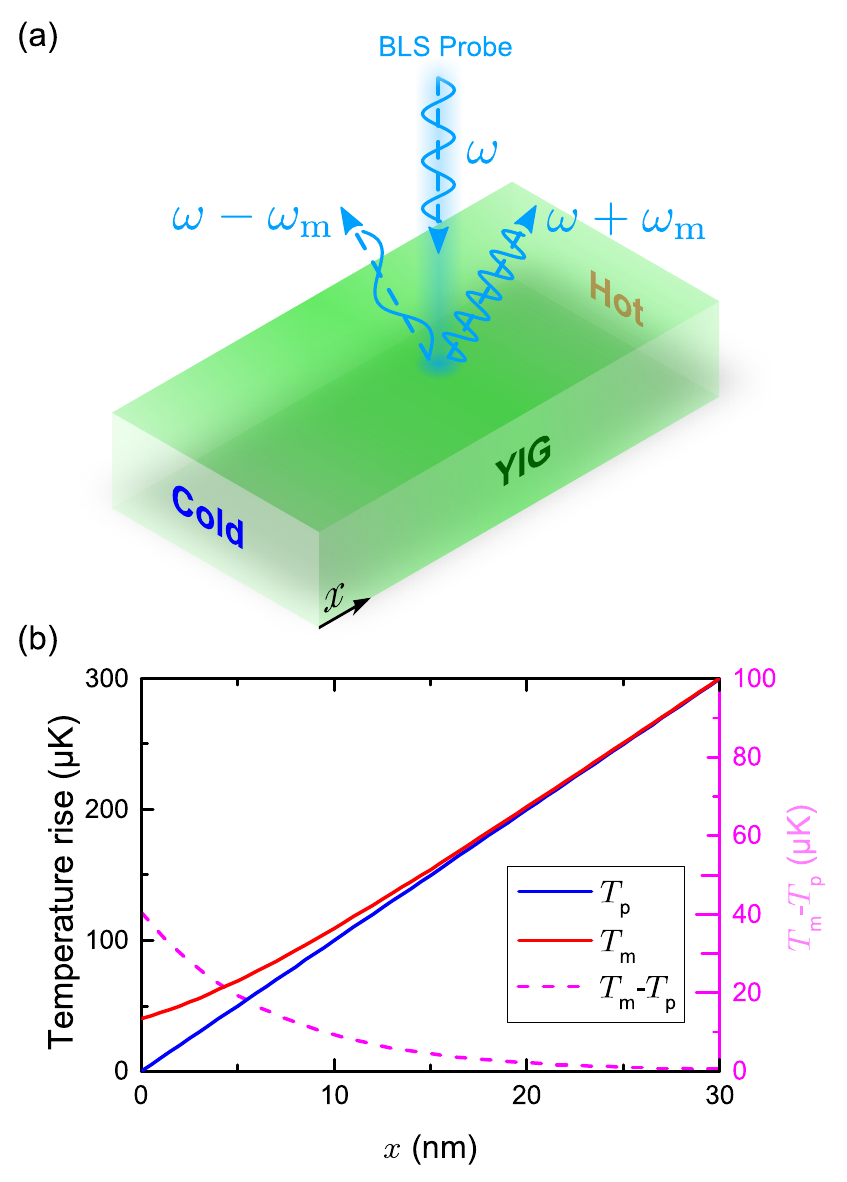}%
\caption{(Color online) (a) Agrawal \textit{et al.}.~\cite{Agrawal2012} probed the magnon temperature in a YIG film along a longitudinal thermal gradient by Brillouin light scattering (BLS). In the BLS experiment light of frequency $\omega$ is scattered inelastically at the magnons in the YIG and reflected back to a detector. The change in frequency $\omega_\mathrm{m}$ is then related to the magnons' temperature.~\cite{Demokritov2008} (b) Phonon and magnon temperature profiles calculated as detailed in Sec.~\ref{sec:analytic} for the first $\SI{2}{\micro m}$ of the $\SI{10}{\milli m}$ long YIG film ($T_\mathrm{R}-T_\mathrm{L}=\SI{100}{K}$) as investigated by Agrawal \textit{et al.}.~\cite{Agrawal2012} One can see that only very close to the sample end at $x=0$ (and $x=\SI{10}{\milli m}$, not shown) a substantial temperature difference $T_\mathrm{m}-T_\mathrm{p}$ arises which is, however, still smaller than the experimental temperature stability of $\pm\SI{0.3}{K}$.}%
\label{fig:MagTempAgrawal}%
\end{figure}

\begin{figure}%
\includegraphics[width=\columnwidth]{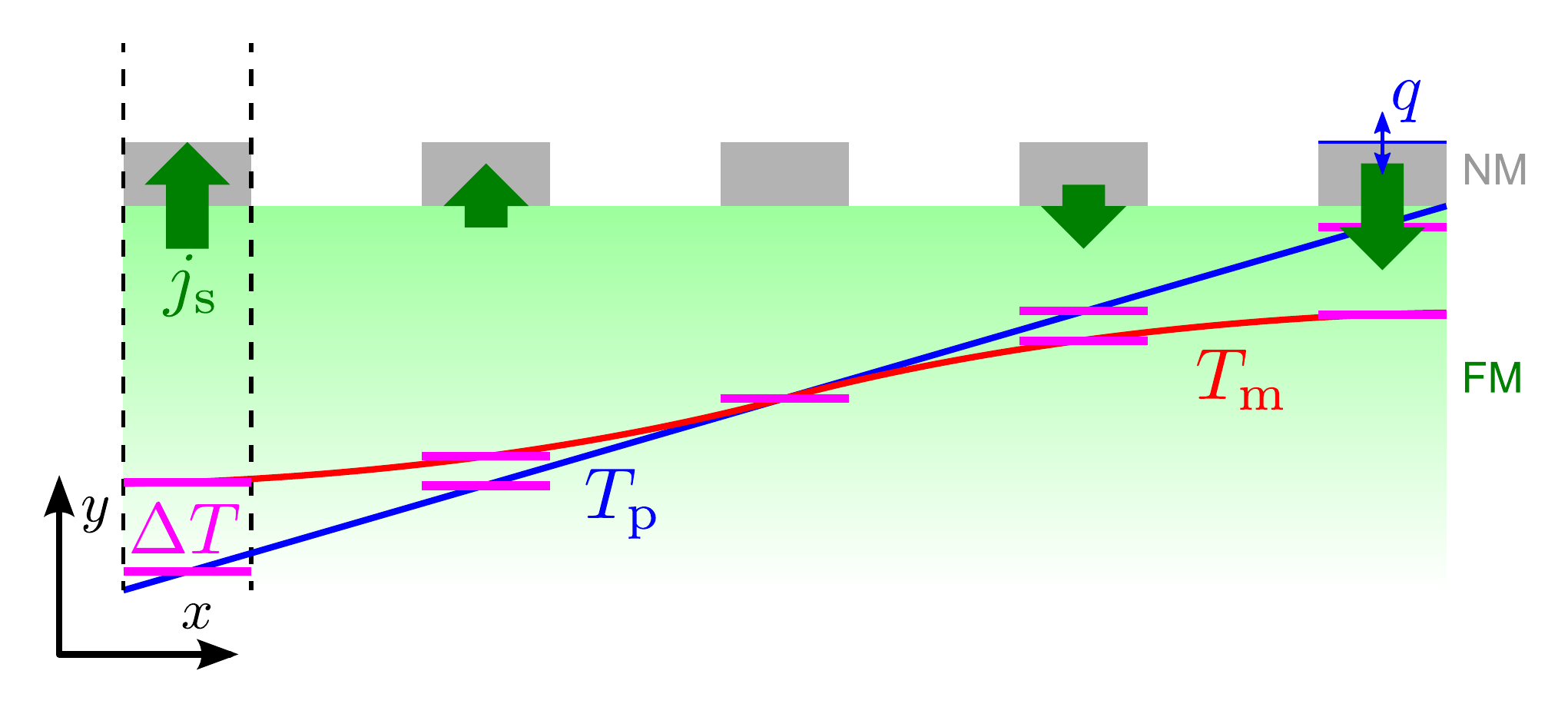}%
\caption{(Color online) Schematic depiction of the magnon and phonon temperature profiles in the transverse spin Seebeck geometry. Here an in-plane ($x$-direction) thermal gradient in the ferromagnet (FM) causes $\dT\approx\Delta T_\mathrm{mp}=T_\mathrm{m}-T_\mathrm{p}$ to vary along the length of the sample, which is reflected in the magnitude and sign of the spin current injected into the normal metal (NM). The signature of the transverse spin Seebeck effect is the sign reversal of the injected spin current and thus the measured voltage around the center of a sample. The platinum layer may, however, be thermally coupled to an external reservoir via a heat current $q$ which will then induce out-of-plane thermal gradients ($y$-direction) that can cause contributions from the longitudinal spin Seebeck effect. Since our calculations suggest that $\Delta T_\mathrm{mp}$ is extremely small in millimeter sized samples which are typically used in experiment, this contribution can become dominant.}%
\label{fig:transSSE}%
\end{figure}

Finally, we also address the temperature profiles in the transverse spin Seebeck effect measurement geometry, in which the externally applied thermal gradient and the emerging spin current are orthogonal to each other.~\cite{Uchida2008} 
\\
Agrawal \textit{et al.}~\cite{Agrawal2012} performed Brillouin light scattering (BLS) experiments [Fig.~\ref{fig:MagTempAgrawal}(a)] on a $\SI{3}{\milli m}\times\SI{10}{\milli m}\times\SI{6.7}{\micro m}$ YIG film (without normal metal stripes on top) in which they measured the magnon and phonon temperature along the direction of an applied thermal gradient.
Based on their data these authors conclude that magnons and phonons have almost identical temperatures, as no systematic difference between $T_\mathrm{m}$ and $T_\mathrm{p}$ could be resolved in the BLS experiments. Figure~\ref{fig:MagTempAgrawal}(b) shows that by applying Eqs.~\eqref{eq:Tm} and~\eqref{eq:Tp} to a sample in this geometry
this is a natural result: From the solution of the 1D heat transport equations, one would not expect a difference between $T_\mathrm{m}$ and $T_\mathrm{p}$ large enough to be detectable by BLS experiments. Using the material parameters for YIG and $T_\mathrm{R}-T_\mathrm{L}=\SI{100}{K}$ as in the experiment by Agrawal \textit{et al.}, our modeling shows that the temperature difference between the magnons and the phonons becomes substantial only very close to the edges of the sample, with $\Delta T_\mathrm{mp}\leq\SI{20}{\micro K}$. This temperature difference is substantially smaller than the temperature stability of $\pm\SI{0.3}{K}$ quoted by Agrawal \textit{et al.}. Furthermore, according to our calculation, $\Delta T_\mathrm{mp}$ is reasonably large only over a length of about $\SI{20}{\nano m}$ which is much less than the lateral resolution ($\SI{40}{\micro m}$) of the experiment. Our calculations thus corroborate the experimental observation that $T_\mathrm{m}\cong T_\mathrm{p}$ in this geometry. These results do not change qualitatively if the macrospin model is used. Turning the argument around, the agreement with the experiment supports the calculations presented in this paper.\\
The fact that no substantial $\Delta T_\mathrm{mp}$ can arise in large samples has an important implication. It means that our simulations fail to reproduce the observed \textit{transverse} spin Seebeck effect~\cite{Uchida2008,Uchida2010a,Jaworski2010} (cf. Fig.~\ref{fig:transSSE}). Especially towards the center of a sample the Sanders and Walton approach, i.e., the solution of the heat transport problem, gives extremely small temperature differences $\dT$, such that great care has to be taken to exclude any spurious contributions to the measured voltages.~\cite{Uchida2008,Uchida2010a} Thermal short circuiting of the platinum on top of the YIG, e.g. via the electrical contacts/wire bonds, black body radiation, or heat transfer to the atmosphere, may introduce thermal gradients normal to the films,~\cite{Huang2011, Avery2012} which in combination with the interfacial thermal resistance can result in major contributions from the \textit{longitudinal} spin Seebeck effect as discussed above.\\
In an attempt to model this problem we set up the geometry and boundary conditions detailed in Ref.~\onlinecite{Uchida2012} (a $\SI{8}{\milli m}\times\SI{4}{\milli m}\times\SI{3.9}{\micro m}$ La:YIG film~\footnote{Following Fig.~7(a) in Ref.~\onlinecite{Uchida2012} we assume a substarte length of $\SI{10}{\milli m}$ and contact area with the heater and heat sink at each end of $\SI{1}{\milli m}\times\SI{4}{\milli m}$, as shown in the same reference in Fig.~6(a). Furthermore, we use the material parameters of YIG for the La:YIG film.} with $\SI{15}{\nano m}$ thick, $\SI{100}{\micro m}$ wide platinum stripes distributed across the La:YIG film) in the 3D FEM and calculate the coupled magnon, electron, and phonon temperature distributions. 
The simulation result suggests that a mean temperature difference at the YIG/Pt interface between the magnons in the ferromagnet and the electrons in the normal metal of just $|\dT|\lesssim\SI{1e-6}{K}$ (at $T_\mathrm{L}-T_\mathrm{R}=\SI{20}{K}$) on the outmost platinum stripe in the case of absent thermal short circuiting ($q=0$ in Fig.~\ref{fig:transSSE}) of the platinum layer will arise. If phonon and electron temperature at the top of the platinum layer are changed by just $\SI{1}{K}$ ($q\neq0$ in Fig.~\ref{fig:transSSE}) from the equilibrium temperature in the previous case one gets $|\dT|\approx\SI{1.5e-5}{K}$, much larger than before. Our simulations therefore suggest that in this measurement geometry, one has to take extreme care to exclude, or at least contain parasitic out-of-plane thermal gradients to a very small level to resolve the transverse spin Seebeck effect.

\section{Conclusion}
We computed the magnon, phonon and electron temperature profiles in typical thin film samples used for spin Seebeck experiments and compared results with experimental data. 
Starting from a 1D analytical diffusion model we found that the thermal coupling between the magnons in the ferromagnet and the electrons in the normal metal notably affects the magnon temperature in the ferromagnet. A Macrospin and thermal magnon model were compared and it was shown that both yield similar spin currents across the ferromagnet/normal metal interface. Also, good agreement between the spin Seebeck effect voltages observed in a series of YIG/Pt samples and the theory by Xiao \textit{et al.}~\cite{Xiao2010} using 3D finite element simulations was found. We also calculated the transverse spin Seebeck effect and showed that, in contrast to earlier predictions~\cite{Xiao2010} and despite the relatively weak coupling between microwave magnons and phonons, magnon and phonon temperature differences were very small as observed in recent experiments.~\cite{Agrawal2012} In contrast to several experiments we therefore do not find a significant transverse spin Seebeck effect in our model. We conclude that the effect is caused by effects beyond the simple diffusion model, such as coherent~\cite{Adachi2013} or subthermal~\cite{Tikhonov2013} phonon coupling through the sample or the substrate.\\

We would like to thank S. Meyer, M. Althammer, M. Opel and S. Gepr\"ags for their help in sample fabrication and T. Brenninger for technical support. Financial support from the DFG via SPP 1538 ``Spin Caloric Transport'', Project No. GO 944/4-1, BA 2954/1-1, FOM (Stichting voor Fundamenteel Onderzoek der Materie), EU-ICT-7 ``MACALO'', the ICC-IMR, Grand-in-Aid for Scientific Research (Kakenhi) A 25247056 and the German Excellence Initiative via the Nanosystems Initiative Munich (NIM) is gratefully acknowledged.

\appendix
\section{Kapitza Resistance}\label{sec:Kapitza}
This Appendix addresses the interfacial thermal resistance (also referred to as thermal contact or Kapitza resistance~\cite{Kapitza1941}) which stems from the scattering of the heat carriers at an interface. The interfacial thermal resistance has been measured for a large number of interfaces,~\cite{Costescu2003,Gundrum2005} however, to the best of our knowledge, no experimental data are available for the YIG/Pt interface. We will therefore utilize established models to calculate the interfacial thermal resistance in our samples.\\
The heat flow $q$ across an interface can be expressed in linear response (``Ohm's law'') by:
\begin{equation}
	q=\frac{1}{R_\mathrm{th}}\Delta T.\label{eq:qgen}
\end{equation}
If the transmission probability of the heat carriers across the interface is $\Gamma$, the associated heat flow $q$, i.e. the amount of energy $U$ transported across the interface per unit area $A$ and unit time $\delta t$ can be written as
\begin{equation}
	q=\frac{U}{A\delta t}\Gamma\approx\frac{\pfrac{U}{T}\Delta T}{A\delta t}\Gamma.
\label{eq:qgen2}
\end{equation}
Combining Eqs.~\eqref{eq:qgen} and~\eqref{eq:qgen2} yields
\begin{eqnarray}
	\left(R_\mathrm{th}\right)^{-1}	&=&	\pfrac{U}{T}\frac{1}{A\delta t}\Gamma\notag\\
	&=&	C\frac{l}{V}\frac{1}{\delta t}\Gamma\notag\\
	&=&	\frac{C}{V}v_\mathrm{g}\Gamma,
\label{eq:Rthgen2}
\end{eqnarray}
where $C=\pfrac{U}{T}$ is the heat capacity and $v_\mathrm{g}=\frac{l}{\delta t}$ is the group velocity of the heat carriers.\\
For phonons, we have to use the phonon group velocity $v_\mathrm{ph}$ and the heat capacity per unit volume $c=C/V$ is calculated for each acoustic branch $j$:
\begin{equation}
	c_j=\frac{C_j}{V}=\dfrac{}{T}\int_0^\infty{\hbar\omega D_j(\omega)n(\omega,T)d\omega},
\label{eq:c}
\end{equation}
where $D_j(\omega)$ is the phonon density of states and $n(\omega,T)$ is the Bose-Einstein distribution function. For the transmission probability $\Gamma$ two models are generally used. The \textit{acoustic mismatch model}~\cite{little1959} (AMM) assumes that the phonons are scattered according to Snell's law at the interface while the \textit{diffusive mismatch model}~\cite{swartz1987} assumes diffuse scattering. 
In the following, we adopt the acoustic mismatch model since all of our interfaces have been grown epitaxially and can be considered flat on a lengthscale corresponding to the wavelength of the relevant acoustic phonons. Anyway, the interfacial thermal resistance obtained from the acoustic and diffusive mismatch models agree with each other within one order of magnitude for all interfaces examined in this paper such that choosing one over the other should not significantly alter the results presented here.\\
The interfacial thermal resistance in the acoustic mismatch model reads:~\cite{swartz1989}
\begin{eqnarray}
\left(R_\mathrm{th}^\mathrm{p}\right)^{-1}&=&\frac12\sum_j v_{1,j}\Gamma_{1,j}\notag\\
&&\times\!\!\int_0^\infty{\hbar\omega\dfrac{\left[D_{1,j}(\omega)n(\omega,T)\right]}{T}d\omega},\label{eq:AMM}\\
	\Gamma_{1,j}&=&\int_0^{\pi/2}{\alpha_{1\to2}^\mathrm{AMM}(\theta,j)\cos\theta\sin\theta d\theta},\\
	 \alpha_{1\to2}^\mathrm{AMM}(\theta_1,j)&=&\frac{\frac{4\varrho_2v_{2,j}}{\varrho_1v_{1,j}}\cdot\frac{\cos\theta_{2,j}}{\cos\theta_{1,j}}}{\left(\frac{\varrho_2v_{2,j}}{\varrho_1v_{1,j}}+\frac{\cos\theta_{2,j}}{\cos\theta_{1,j}}\right)^2},\label{eq:alpha}\\
	\notag
\end{eqnarray}
where $\theta_{2,j}$ is linked to $\theta_{1,j}$ (the angle of the outgoing and incident phonons) by Snell's law of acoustic waves~\cite{rayleigh1894}
\begin{equation}
	v_{2,j}\sin\theta_{1,j}=v_{1,j}\sin\theta_{2,j},
	\label{eq:Snell}
\end{equation}
where $v_{i,j}$ is the speed of sound, and $j\in\{1,2,3\}$ denotes the pressure ($j=1$) and shear wave ($j=2,3$) phonon branches. The index $i\in\{1,2\}$ denotes the materials on the left and right side of an interface. The full expression for $\alpha_{1\to2}^\mathrm{AMM}(\theta_1,j)$ in Eq.~\eqref{eq:alpha} was adopted from Ref.~\onlinecite{little1959}.
We calculate the interfacial thermal resistance at $T=\SI{300}{K}$ in the Debye approximation~\cite{Debye1912} and obtain the Debye frequencies $\omega_{\mathrm{c},i,j}$ from the longitudinal and transverse speeds of sound $v_{i,j}$ by~\cite{Hunklinger2005}
\begin{equation}
	\omega_{\mathrm{c},i,j}=\left(6\pi n_i\right)^\frac{1}{3}v_{i,j},
\label{eq:cutoff}
\end{equation}
where $n$ is the atomic density of the material. The Debye model is a good approximation for simple crystal structures and should be appropriate for the long wavelength phonons in (cubic) YIG, but is too crude to accurately describe its complex phonon dispersion at large wave vectors.\\
Using Eq.~\eqref{eq:AMM} in the Debye approximation and the material parameters summarized in Table~\ref{tab:matpar}, we arrive at values of $(R_\mathrm{th}^\mathrm{p,Pt/YIG})^{-1}=\SI{2.79e8}{W/(m^2\,K)}$ for the YIG/Pt interface, $(R_\mathrm{th}^\mathrm{p,YIG/GGG})^{-1}=\SI{2.04e8}{W/(m^2\,K)}$ for the YIG/GGG interface and $(R_\mathrm{th}^\mathrm{p,YIG/YAG})^{-1}=\SI{1.27e8}{W/(m^2\,K)}$ for the YIG/YAG interface, respectively. These results agree well with experimental data obtained for similar interfaces.~\cite{Costescu2003,Gundrum2005}\\
In addition to the YIG/Pt heterostructures, we also investigated samples with an additional metallic (gold) buffer layer between the platinum and the YIG. This introduces an additional metal/metal interface at which the thermal transport is dominated by the electrons. Following Ref.~\onlinecite{Drchal2002}, the majority of electrons scatter diffusively at the interface since the Fermi wavelength is in the \AA ngstr\"{o}m regime and therefore smaller than the typical interface roughness even for very smooth interfaces in heteroepitaxial composites. Therefore the diffusive mismatch model is modified to account for the electronic transport:~\cite{Gundrum2005}
\begin{eqnarray}
		(R_\mathrm{th}^\mathrm{e})^{-1}	&=&	\frac12v_1(E_\mathrm{F})\Gamma_1(E_\mathrm{F})\notag\\
		&&\times\int_0^\infty{E\dfrac{\left[D_1(E)n(E,T)\right]}{T}dE}\label{eq:Rel},\\
			\Gamma_1(E)	&=&	\int_0^{\frac\pi2}{\frac{v_2(E)D_2(E)\cos\theta\sin\theta}{v_1(E)D_1(E)+v_2(E)D_2(E)} d\theta},
\end{eqnarray}
with $D_i(E)$ and $n(E,T)$ the electronic density of states in the material $i$ and Fermi-Dirac distribution function, respectively. $v_1$ and $v_2$ are the electron velocities on both sides of the interface and $E_\mathrm{F}$ is the Fermi energy. The integral in Eq.~\eqref{eq:Rel} coincides with the one for the electronic heat capacity $C_\mathrm{e}$, which for a degenerate electron gas is $C_\mathrm{e}=(\pi^2/3)D(E_\mathrm{F})k_\mathrm{B}^2T=\gamma_\mathrm{S} T$ with the Sommerfeld constant $\gamma_\mathrm{S}$. Thus, $R_\mathrm{th}^\mathrm{e}$ can be written as
\begin{eqnarray}
	\left(R_\mathrm{th}^\mathrm{e}\right)^{-1}&=&\frac{Z_1Z_2}{4(Z_1+Z_2)},\label{eq:DMMel}\\
	Z_i&=&\gamma_{\mathrm{S},i} v_{\mathrm{F},i}T,
\end{eqnarray}
where $v_{\mathrm{F},i}$ is the Fermi velocity in the material $i$. For the platinum/gold interface ($\gamma_\mathrm{S}^\mathrm{Pt}=\SI{748.1}{J/(m^3\,K^2)}$,~\cite{Lin2008} $v_\mathrm{F}^\mathrm{Pt}=\SI{2.19e5}{m/s}$,~\cite{Ketterson1968} $\gamma_\mathrm{S}^\mathrm{Au}=\SI{67.6}{J/(m^3\,K^2)}$,~\cite{Lin2008} $v_\mathrm{F}^\mathrm{Au}=\SI{1.0e5}{m/s}$~\cite{Lengeler1977}) we obtain a contribution from the electrons [$(R_\mathrm{th}^\mathrm{e,Pt/Au})^{-1}=\SI{3.691e9}{W/(m^2\,K)}$] which is notably larger than the contribution from the phonons [$(R_\mathrm{th}^\mathrm{p,Pt/Au})^{-1}=\SI{1.325e9}{W/(m^2\,K)},\ (R_\mathrm{th}^\mathrm{p,Au/YIG})^{-1}=\SI{1.63e8}{W/(m^2\,K)}$], in good agreement with experimental results.~\cite{Gundrum2005}\\

\section{Magnon-phonon relaxation time}\label{sec:tmp}
Let $\bm{m}$ be the unit vector parallel to the magnetization precessing around the $\hat{\bm{z}}$-axis. Following Ref.~\onlinecite{Xiao2010}, the magnon temperature $T_\mathrm{m}$ may then be parameterized by the thermal suppression of the average magnetization $\propto(1-\langle m_z(t)\rangle)$ with $\langle\cdot\rangle$ denoting the ensemble average.
Since each magnon decreases $m_z$ by $\hbar$, $\langle m_z(t)\rangle$ measures the total number of magnons $N=M_\mathrm{s}V(1-\langle m_z\rangle)/(\gamma\hbar)$ in a volume $V$. With
\begin{eqnarray}
	N &=& V\int{\frac{4\pi k^2}{(2\pi)^3}\frac{1}{\mathrm{e}^{\beta\hbar\omega_k}-1}dk}\notag\\
	&=& \frac{V}{8\pi^\frac{3}{2}}\left(\frac{k_\mathrm{B}T_\mathrm{m}}{D}\right)^\frac{3}{2}\mathrm{Li}_{3/2}\left(\mathrm{e}^{\beta\hbar\omega_0}\right)\notag\\
	&\approx& \frac{V\zeta(3/2)}{8\pi^\frac{3}{2}}\left(\frac{k_\mathrm{B}T_\mathrm{m}}{D}\right)^\frac{3}{2},
	\label{eq:Nmag}
\end{eqnarray}
where $\beta=1/(k_\mathrm{B}T_\mathrm{m})$, $\hbar\omega_k=\hbar\omega_0+Dk^2$ and the ferromagnetic resonance frequency given by $\hbar\omega_0\approx\hbar\gamma\mu_0\sqrt{H(H+M_\mathrm{s})}\ll k_\mathrm{B}T $ (Ref.~\onlinecite{Kittel1948}, $\mu_0H$ being the externally applied in-plane magnetic field and $\mu_0$ the vacuum permeability) we have
\begin{eqnarray}
	\frac{d}{dt}(1-\langle m_z\rangle) &=& \frac{\gamma\hbar}{M_\mathrm{s}V}\frac{dN}{dt}\notag\\
	&=& \frac{\gamma\hbar}{M_\mathrm{s}V}\frac{d}{dt}\left[\frac{V\zeta(3/2)}{8\pi^\frac{3}{2}}\left(\frac{k_\mathrm{B}T_\mathrm{m}}{D}\right)^\frac{3}{2}\right]\notag\\
	&=& \frac{\gamma\hbar}{M_\mathrm{s}}\frac{3\zeta(3/2)}{16\pi^\frac{3}{2}}\left(\frac{k_\mathrm{B}}{D}\right)^\frac{3}{2}T_\mathrm{m}^\frac12\frac{dT_\mathrm{m}}{dt}.\label{eq:dmdt1}
\end{eqnarray}
In YIG there is no damping by electrons, hence Eq.~(D8) in Ref.~\onlinecite{Xiao2010} reads
\begin{equation}
	\frac{d\langle m_z\rangle}{dt}	=	\frac{\alpha_\mathrm{G}}{1+\alpha_\mathrm{G}^2}\frac{2\gamma k_\mathrm{B}}{M_\mathrm{s}V_\mathrm{a}}(T_\mathrm{m}-T_\mathrm{p}).
\label{eq:dmdt2}
\end{equation}
Equating Eq.~\eqref{eq:dmdt1} and Eq.~\eqref{eq:dmdt2}, we have
\begin{eqnarray}
	\frac{dT_\mathrm{m}}{dt}	&=&	-\frac{\alpha_\mathrm{G}}{1+\alpha_\mathrm{G}^2}\frac{2\zeta(5/2)}{\zeta(3/2)}(T_\mathrm{m}-T_\mathrm{p})\notag\\
	&\approx&	-1.03\alpha_\mathrm{G}\frac{k_\mathrm{B}T_\mathrm{m}}{\hbar}(T_\mathrm{m}-T_\mathrm{p})
\end{eqnarray}
and a comparison with Eq.~\eqref{eq:mptime} then yields
\begin{equation}
	\tau_\mathrm{mp}=\frac{c_p}{c_t}\frac{\hbar}{1.03\alpha_\mathrm{G}k_\mathrm{B}T_\mathrm{m}}\approx\frac{\hbar}{\alpha_\mathrm{G}k_\mathrm{B}T_\mathrm{m}}.
\end{equation}

\bibliography{Magnon_phonon_and_electron_temperature_profiles_and_the_spin_Seebeck_effect_in_magnetic_insulator_normal_metal_hybrid_structures}
\end{document}